\documentclass[12pt]{iopart}
\usepackage{graphicx}
\usepackage{amssymb}
\usepackage[english]{babel}
\usepackage{psfrag}
\newcommand{\bq}{\begin{equation}}
\newcommand{\eq}{\end{equation}}
\newcommand{\ba}{\begin{eqnarray}}
\newcommand{\ea}{\end{eqnarray}}
\usepackage{color}

\begin{document}

\title{Stochastic Description of a Bistable Frustrated Unit}
\author{Ashok Garai$^1$, Bartlomiej Waclaw$^2$, Hannes Nagel$^3$, and Hildegard Meyer-Ortmanns$^1$}
\address{$^1$School of Engineering and Science, Jacobs University Bremen, P.O.Box 750561, D-28725 Bremen, Germany}
\address{$^2$School of Physics and Astronomy, University of Edinburgh, Mayfield Road, Edinburgh EH9 3JZ, United Kingdom}
\address{$^3$Institut f\"ur Theoretische Physik, Universit\"at Leipzig,\\Postfach 100\,920, 04009 Leipzig, Germany}
\eads{\mailto{ashok.garai@gmail.com}, \mailto{bwaclaw@staffmail.ed.ac.uk}, \mailto{hannes.nagel@itp.uni-leipzig.de}, \mailto{h.ortmanns@jacobs-university.de}}

\begin{abstract}
Mixed positive and negative feedback loops are often found in biological systems which support oscillations.
In this work we consider a prototype of such systems, which has been recently found at the core of many genetic circuits showing oscillatory behaviour.
Our model consists of two interacting species A and B, where A activates not only its own production, but also that of its repressor B. While the self-activation of A leads already to a bistable unit, the coupling with a negative feedback loop via B makes the unit frustrated. In the deterministic limit of infinitely many molecules, such a bistable frustrated unit is known to show excitable and oscillatory dynamics, depending on the maximum production rate of A which acts as a control parameter. We study this model in its fully stochastic version and we find oscillations even for parameters which in the deterministic limit are deeply in the fixed-point regime. The deeper we go into this regime, the more irregular these oscillations are, becoming finally random excitations whenever fluctuations allow the system to overcome the barrier for a large excursion in phase space. The fluctuations can no longer be fully treated as a perturbation. The smaller the system size (the number of molecules), the more frequent are these excitations. Therefore, stochasticity caused by demographic noise makes this unit even more flexible with respect to its oscillatory behaviour. \end{abstract}


\maketitle

\section{Introduction}
\label{intro_sec}
Biological systems such as populations of interacting organisms or genetic networks are inherently non-linear due to feedback loops and stochastic due to a finite number of contributing ingredients and a discrete set of reaction events. One of the various origins for stochastic behaviour is demographic noise caused by fluctuations in the population size. Although often ignored in the past, it is now well established that such inherent stochasticity is not just a small correction to deterministic, infinite-population solutions, but it may lead to a number of qualitatively new effects. As already demonstrated, demographic noise can lead to temporal oscillations for predator-prey models \cite{newman}, counter-intuitive reversions in the evolutionary process for replicator dynamics \cite{traulsen}, or persistent spatial patterns \cite{golden1}. In gene expression models, it has been shown for a single self-regulating gene that the solution of the stationary state reveals qualitative deviations from the deterministic solution like anticooperative behaviour in the limit of slow binding and unbinding rates \cite{alex1,alex2}. In networks of mutually interacting genes, extinction and resurrection events for small population size can lead to additional fixed points in the attractor landscape \cite{alex3}. In the repressilator model, a prototype of genetic oscillators, coherence resonance has been observed due to stochastic fluctuations \cite{yoda}, and a modification of frequency, amplitude and parameter regime where oscillations occur has been demonstrated~\cite{loinger}. Similar effects of intrinsic noise have been seen in stochastic delay systems of genes \cite{galla1}.

The aim of this work is to analyse the effect of finite-population demographic noise on a so-called bistable frustrated unit (BFU). This model consists of two interacting species A,B. Species A activates its own production and without coupling to the second species B, the self-loop of A would correspond to a bistable system. But A here also activates the production of its repressor B. The unit of the two loops was then termed frustrated in analogy to antiferromagnetic frustrated couplings \cite{sandeep}. The reason is that A gets conflicting input, activation from itself and repression via the activation of the second species B. Throughout this paper we use ``bistable frustrated unit" as a name for our model. What makes this simple model particularly interesting is that by tuning the ratio of half life of the two species, one can obtain fast dynamics for the activating species A and slow dynamics for the repressing species B, similarly to the fast and slow degrees of freedom of FitzHugh-Nagumo units as models for neural networks \cite{fitzhugh,nagumo}. This defines an intrinsic ratio of time scales with implications for the shape of amplitudes, the form of limit cycles, and the probability distribution in phase space.

The deterministic (infinite-population) version of this model was originally proposed in Ref.~\cite{sandeep} as an effective, coarse-grained description of genetic circuits in which a system, bistable due to a positive feedback loop, is coupled to a negative feedback loop which adds frustration to the system. The model has been shown to have a variety of oscillatory behaviour, which is easily tunable by the model parameters.
In addition to the limit-cycle oscillatory regime, two fixed-point regimes have been recently identified  in this model \cite{hmokaluza}.

In this paper we shall analyse a fully stochastic version of this model with a finite number of particles of each species.
We shall see that, in contrast to the deterministic limit, cycles (, i.e.,oscillations) appear even if the parameters of the model are deeply in the deterministic fixed-point regime. The existence of such ``quasi-cycles'', as we shall call them in order to stress their difference to usual, deterministic cycles, indicates that the range of parameters for which the stochastic system exhibits oscillatory behaviour is much broader, as argued in Ref.~\cite{hwa} for a similar system. This suggests that fine-tuning of the parameters is less critical.
This observation is relevant for real-world genetic circuits when they are realized as single units and exhibit oscillations such as the circadian clock \cite{leibler}, because it means that a finite number of particles can make them more robust to damage, e.g., caused by mutations changing protein production rates, binding affinities etc. The cost to pay is, as we shall see, less regular oscillations.

A lower sensitivity to the very parameter choice is also seen in a generic model of pattern formation \cite{golden2}, where it has been demonstrated that so called ``quasi-patterns'', very similar to Turing patterns \cite{turing}, occur even without fine-tuning of the parameters. Turing patterns are observed in a variety of systems, so along with them also quasi-patterns may be observed in very different applications, ranging from ecological systems as studied in \cite{golden1,golden2} to populations of players, playing a kind of prisoner's dilemma game \cite{galla0} to neural networks \cite{bressloff} to genetic networks \cite{yoda,loinger,galla1}.

Similar observations have been made for systems more closely related to our unit. For example, quasi-cycles in the stochastic brusselator model \cite{galla2} were detected well away from the regime of oscillations in the deterministic model. However, in contrast to these recent findings which could be well understood by treating demographic noise as a perturbation (although often amplified by stochastic resonance) on top of the deterministic behaviour, we shall see that noise has more dramatic effects on our system, going beyond the perturbative regime.

The paper is organized as follows. In section \ref{sec2} we present the deterministic and the stochastic formulations of the model.
Section \ref{sec3} deals with results of Gillespie simulations over different parameter ranges and as function of the system size. We show that quasi-cycles can be distinguished from limit cycles via a distinct, more rapid decay of their autocorrelation function, but apart from that they share other common features, particularly in the transition region. In section \ref{sec4} we shall solve the master equation approximately by using the van Kampen expansion in $1/N_0$, where $N_0$ stands for the system size. We derive the variances, autocorrelation functions and the power spectrum for the fluctuations of the two species, both in the fixed-point phase and in the limit-cycle phase. We show that the van Kampen expansion predicts the behaviour of all these quantities quite far from the  transition points, where large excursions in phase space, induced by large fluctuations, are so rare that they can be neglected, but it breaks down close to fixed-point/limit-cycle transitions with large fluctuations and large excursions in phase space, which are characteristic for the excitatory behaviour of our model. This shows that finite-size effects are even more important in our model than in previously considered models of gene expression, and taking only first-order terms of the expansion in $1/N_0$ largely neglects a characteristic feature of the BFU, that is its excitability. Conclusions and an outlook are presented in section \ref{sec5}.

\begin{figure}[ht]
\center
\includegraphics*[width=0.25\columnwidth]{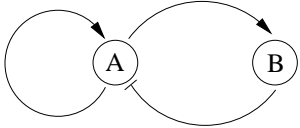}
\caption{Motif of a bistable frustrated unit (BFU). Pointed arrows denote activation (increase in the production rate), the blunt arrow denotes repression (decrease in the production rate).} \label{fig1}
\end{figure}

\section{The model}
\label{sec2}
Let us consider a simple model of frustrated bistability as depicted in Fig.~\ref{fig1}. In this model, A and B are two different species. Although from a statistical physics point of view the identity of A, B is not important, we shall assume here that A, B are two different protein types. This is in line with the original motivation of this model as a coarse-grained description of some genetic circuits. The protein A activates its own production (transcription in the biological language) and also the production of the protein B, which in turn represses the production of A. In this way, we have a self-activating bistable unit coupled to a negative feedback loop. As stated in the introduction, we shall call the motif from  Fig.~\ref{fig1} a bistable frustrated unit (BFU). This motif exists as a part of many genetic circuits, such as the embryonic cell-cycle oscillator \cite{pomerening} and the circadian clock \cite{leibler}.
The simplest, idealised implementation of the BFU has been analysed on a deterministic level, with protein concentrations as the only dynamical variables, and it is known to produce oscillations in a certain range of model parameters \cite{guantes,sandeep}.
In particular, the model studied in Ref.~\cite{sandeep}, which we will adopt as our starting point, assumes that the protein production rates depend on the concentrations $\phi_A$ and $\phi_B$ of the two protein species as follows:
\begin{eqnarray}
 \frac{d\phi_A}{dt}&=&\frac{\alpha}{1+\phi_B/K}\,\frac{b+\phi_A^2}{1+\phi_A^2}-\phi_A, \label{eq-det11} \\
 \frac{d\phi_B}{dt}&=&\gamma(\phi_A-\phi_B), \label{eq-det12}
\end{eqnarray}
where $\gamma$ is the ratio of the half-life of A to that of B. Here we shall focus on the case $\gamma\ll 1$, that is when the protein B has a much longer half-life than A with a slow reaction on changes in A, while A has a fast response to changes in B. The parameter $\gamma$ plays a similar role to the $\epsilon$-parameter in a FitzHugh-Nagumo element that separates the time scales of a fast variable, corresponding to the voltage variable in the context of neurons, and a slow variable, corresponding to the recovery variable there. The parameter $K$ sets the strength of repression of A by B. We shall assume $K\ll 1$, so that already a small concentration of B will inhibit the production of A. The parameter $b$ determines the basal expression level of A. We set this parameter to anything larger than zero but much smaller than one, so that the system cannot be absorbed in the state $\phi_A=\phi_B=0$ and, simultaneously, the production rate of A is small for $\phi_A\approx 0$. In units where the production rate of B is equal to its degradation rate, $K$ plays the role of a Michaelis constant that sets the strength of the repression of A by $K$. As argued in \cite{sandeep}, the choice of the Hill coefficient (here as $h=1$ in $(\phi_B/K)^h$), used in the repression of A by B, is not essential. For the other Hill coefficients (activation of B by A with Hill coefficient $h=0$  and self activation of A involving powers of $h=2$) we made the same choice as in \cite{sandeep,hmokaluza} for better comparability.
The parameter $\alpha$ is the maximal rate of production of A for full activation ($\phi_A^2\gg b$) and no repression ($\phi_B\approx 0$). This will be our tunable parameter which we will use to control the behaviour of our model. This parameter seems to be the easiest one to control in real, experimental systems \cite{sandeep}. In the deterministic description of \cite{hmokaluza}, we extended the parameter range of $\alpha$ independently of the biological relevance. As a function of increasing $\alpha$ we found for small $\alpha$ a fixed-point regime with excitable behavior, followed first by a limit-cycle regime with an unstable fixed point and rich oscillatory behavior in this intermediate range of $\alpha$, and next by a second fixed-point regime with excitable behavior for large $\alpha$. The different regimes (or phases) are separated  by subcritical Hopf bifurcations (for a definition see for example Ref.~\cite{strogatz}) with corresponding hysteresis effects. Typical trajectories in phase space for the three regimes in the deterministic case are shown in Fig.~2 of \cite{hmokaluza} (as well as in Fig.s ~\ref{fig2},~\ref{fig3} below), and the bifurcation diagram is displayed in Fig.~4 of the same reference.

It is not at all obvious that Eqs.~(\ref{eq-det11}--\ref{eq-det12}) can actually be interpreted as coarse-grained description of a full genetic circuit, in which other proteins \cite{ingolia} as well as mRNA \cite{leibler} are typically involved. However, in the forthcoming paper \cite{forthcoming} we shall show that Eqs.~(\ref{eq-det11}--\ref{eq-det12}) can be derived from the assumed underlying reactions between genes, mRNA and proteins in a more detailed model, for a certain range of reaction rate constants.

We now formulate an effective, stochastic counterpart of the model (\ref{eq-det11}--\ref{eq-det12}) by introducing individual molecules of A and B. At each time, the system is characterized by the number of molecules of each species, $N_A$ and $N_B$. The concentrations $\phi_A,\phi_B$ become discrete numbers: $\phi_A=N_A/N_0$ and $\phi_B=N_B/N_0$, where $N_0$ plays the role of the system size. The parameter $N_0$ is not the physical volume, but it allows us to control the average numbers of molecules A and B, which are proportional to $N_0$, and hence to investigate the role of demographic noise in the model. Molecules of both protein species are created and annihilated with certain rates. Since we want to establish the correspondence between the deterministic (\ref{eq-det11}--\ref{eq-det12}) and the stochastic version of the model, we assume the following rates for four possible processes:
\ba
	\mbox{production of A} \qquad & N_A &\to N_A + 1 \quad	\mbox{with rate} \; N_0 f(N_A/N_0,N_B/N_0) \label{eq:rates1}\\
	\mbox{decay of A} \qquad & N_A &\to N_A - 1	\quad \mbox{with rate} \; N_A \\
	\mbox{production of B} \qquad & N_B &\to N_B + 1 \quad	\mbox{with rate} \; \gamma N_A \label{eq:rates3}\\
	\mbox{decay of B} \qquad & N_B &\to N_B - 1	\quad \mbox{with rate} \; \gamma N_B \label{eq:rates4}
\ea
where ``rate'' denotes the transition rate of a specific process, and $f(\phi_A, \phi_B)$ is given by
\begin{equation}
 f(\phi_A, \phi_B) = \frac{\alpha}{1+\phi_B/K}\,\frac{b+\phi_A^2}{1+\phi_A^2}. \label{eq-f}
\end{equation}
It is by no means the only possible way of modelling the BFU on the level of individual molecules and chemical reactions, but we shall see later that this choice of the rates reproduces the deterministic system in the limit $N_0\to\infty$. Also, this is only an effective model which neglects gene states, mRNA concentrations etc., which should be included in a more detailed model \cite{forthcoming}.
Similarly to the deterministic case (\ref{eq-det11}--\ref{eq-det12}), here we focus only on the dynamics of the proteins A, B. It should be noticed that the difference of half-lifes of species A and B here is still implemented via the parameter $\gamma$ in Eqs.~(\ref{eq:rates3}) and (\ref{eq:rates4}), making the effective production and decay rates of B much slower than those of A, without resolving the underlying genetic and mRNA-levels of how these rates are generated.

The time evolution of our system can be described by the following master equation for the probability $P(N_A,N_B)$ for finding $N_A$ proteins of type A and $N_B$ proteins of type B at time $t$:
\begin{eqnarray}
 \frac{\partial P(N_A, N_B)}{\partial t}&=&-(N_0f(N_A/N_0,N_B/N_0)+N_A+\gamma N_A+\gamma N_B)P(N_A, N_B) \nonumber \\
 &+&(N_A+1)P(N_A+1,N_B) \nonumber \\
 &+ &N_0f((N_A-1)/N_0,N_B/N_0)P(N_A-1,N_B) \nonumber \\
&+& \gamma(N_B+1)P(N_A,N_B+1) + \gamma N_A P(N_A, N_B-1).
\label{eq-master}
\end{eqnarray}
On the right-hand-side we have as loss terms for $P(N_A,N_B)$ the production of A with rate $N_0f(N_A/N_0,N_B/N_0)$, the decay of A proportional to $N_A$, the production of B proportional to $\gamma N_A$, and its deletion proportional to $\gamma N_B$. Gain terms to $P(N_A,N_B)$ result from the decay of A out of a state with $N_A+1$ proteins, its production from a state with $N_A-1$ proteins with rate  $N_0f((N_A-1)/N_0,N_B/N_0)$, the decay of B with rate $\gamma (N_B+1)$ and its production with rate $\gamma N_A$ from $N_B-1$ proteins of type B.

The master equation (\ref{eq-master}) is our starting point. In the limit of $N_0\to\infty$, our stochastic description becomes equivalent to the deterministic model (\ref{eq-det11}--\ref{eq-det12}). One can see this by multiplying both sides of Eq.~(\ref{eq-master}) by $N_A$ or $N_B$ and summing over $N_A$ and $N_B$. If we now define the average of some observable $O(N_A,N_B)$ as $\left<O(N_A,N_B)\right>\equiv \sum_{N_A,N_B} O(N_A,N_B)P(N_A,N_B,t)$ and assume that $\langle f(N_A/N_0, N_B/N_0)\rangle= f(\langle N_A\rangle/N_0, \langle N_B\rangle/N_0)$ (which is fulfilled for a sufficiently sharply peaked probability distribution), we obtain the following equations for the averages $\left<N_A\right>,\left<N_B\right>$:
\begin{eqnarray}
 \frac{d\left<N_A\right>}{dt}&=&\frac{\alpha N_0}{1+\frac{\left<N_B\right>}{KN_0}}\frac{b+\left(\frac{\left<N_A\right>}{N_0}\right)^2}{1+\left(\frac{\left<N_A\right>}{N_0}\right)^2}-\left<N_A\right>, \\
 \frac{d\left<N_B\right>}{dt}&=&\gamma(\left<N_A\right>-\left<N_B\right>).
\end{eqnarray}
Substituting $\left<N_A\right>=N_0\phi_A,\left<N_B\right>=N_0\phi_B$ we arrive at our deterministic set of equations (\ref{eq-det11}--\ref{eq-det12}) for the concentrations $\phi_A,\phi_B$.

\section{Computer simulations}
\label{sec3}
The master equation (\ref{eq-master}) cannot be solved exactly due to the strongly non-linear term $f(\phi_A,\phi_B)$. Before we proceed to its solution by approximate methods, we shall discuss the results of computer simulations. We have simulated our stochastic model using the Gillespie algorithm \cite{gillespie}, which samples the exact probability distribution $P(N_A,N_B,t)$ by simulating trajectories $(N_A(t),N_B(t))$ with transition rates (\ref{eq:rates1}--\ref{eq:rates4}). The algorithm actually generates triples of points $\{N_{A,i},N_{B,i},t_i\}$, where $t_i$ is the physical time at $i$-th step of the algorithm\footnote{The physical time $t$ and the computer time $i$ measured in Gillespie steps are not the same, nor are they strictly proportional, because natural time intervals between succeeding reactions are smaller or larger depending on the total rate of all reactions which changes with $N_A,N_B$.}.
In this way, we obtain not only $P(N_A,N_B,t)$ for any finite time $t$, but also autocorrelation functions and power spectra of $N_A(t),N_B(t)$ from the corresponding time series. For example, to obtain $P(N_A,N_B,t)$ from the Gillespie simulations, we determine a histogram that counts how often the trajectory visits a point $(N_A,N_B)$ at a given fixed physical time $t$. The stationary distribution $P^*(N_A,N_B)=P(N_A,N_B,t\to\infty)$ is obtained by simulating the system for some time $t_0$ until it reaches the stationary state, and collecting the histogram of visited points for all $t>t_0$.
Probability distributions, derived from the Gillespie trajectories in this way, agree very well with direct numerical integration of the master equation (\ref{eq-master}).

\begin{figure}
	\includegraphics*[width=0.33\columnwidth]{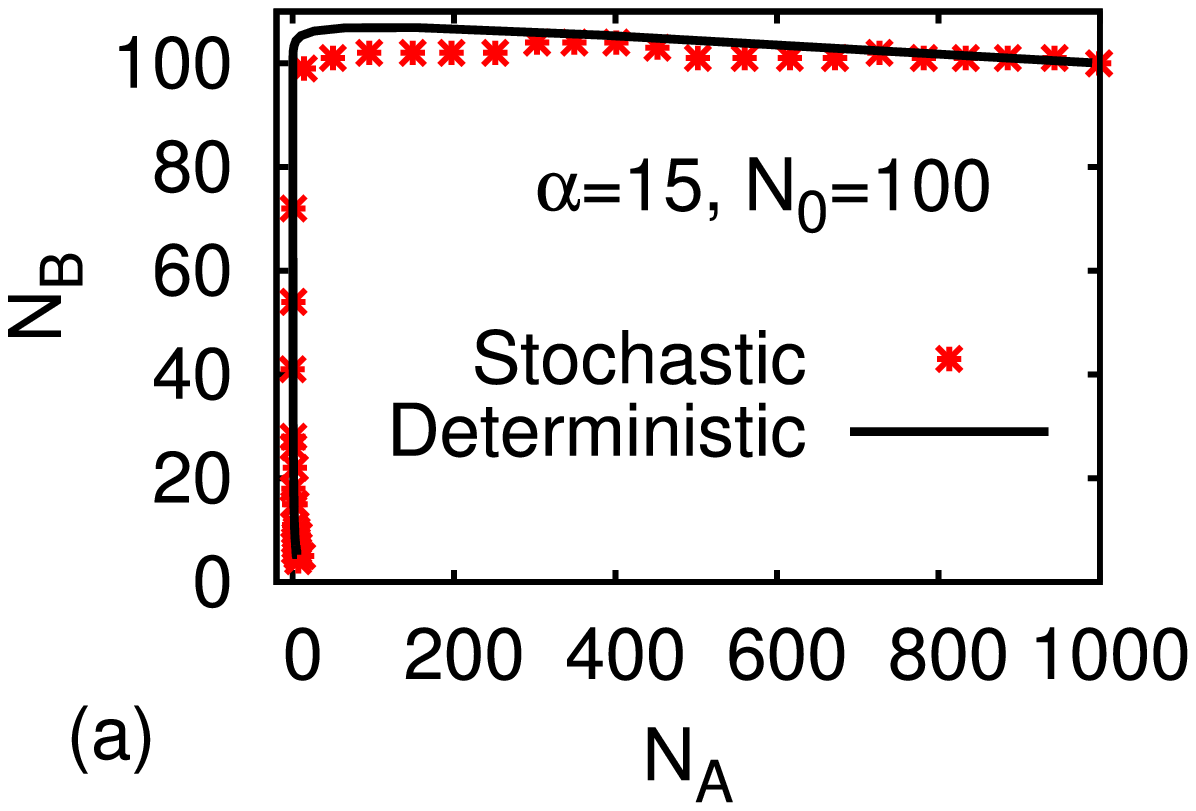}
	\includegraphics*[width=0.33\columnwidth]{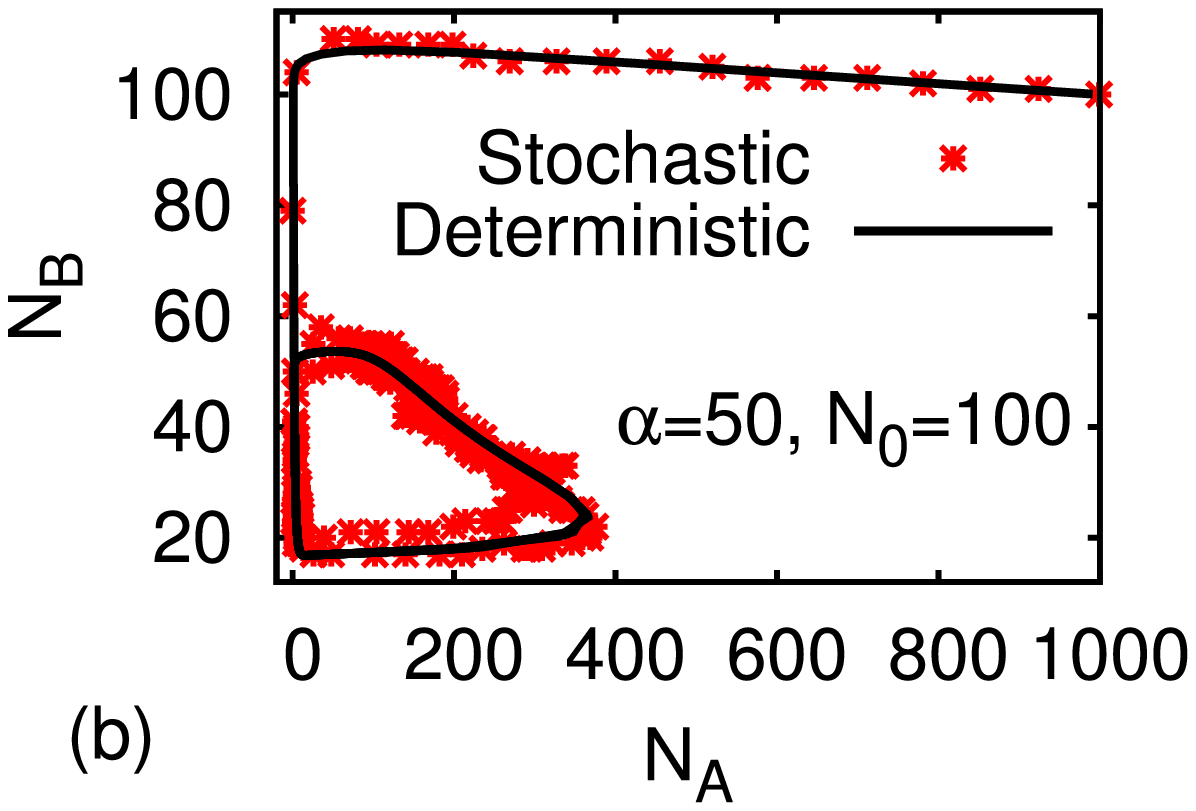}
	\includegraphics*[width=0.33\columnwidth]{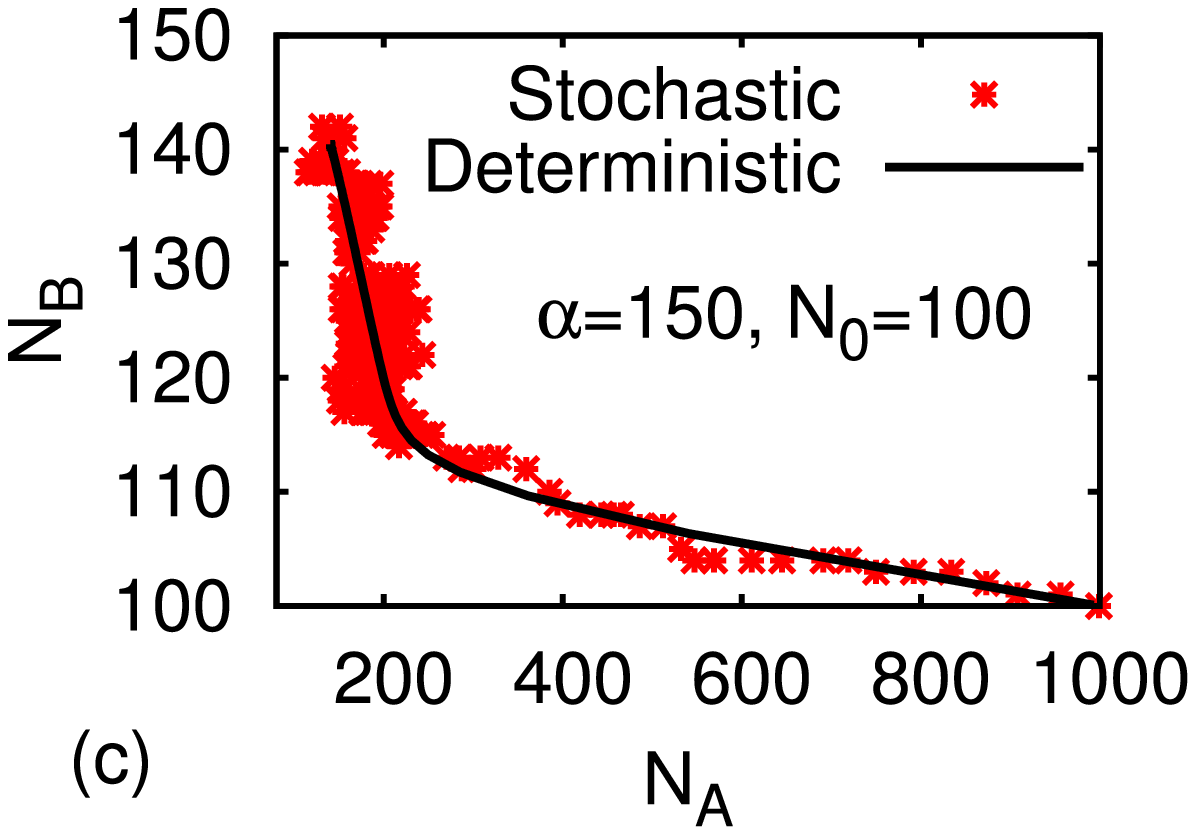}
	\includegraphics*[width=0.33\columnwidth]{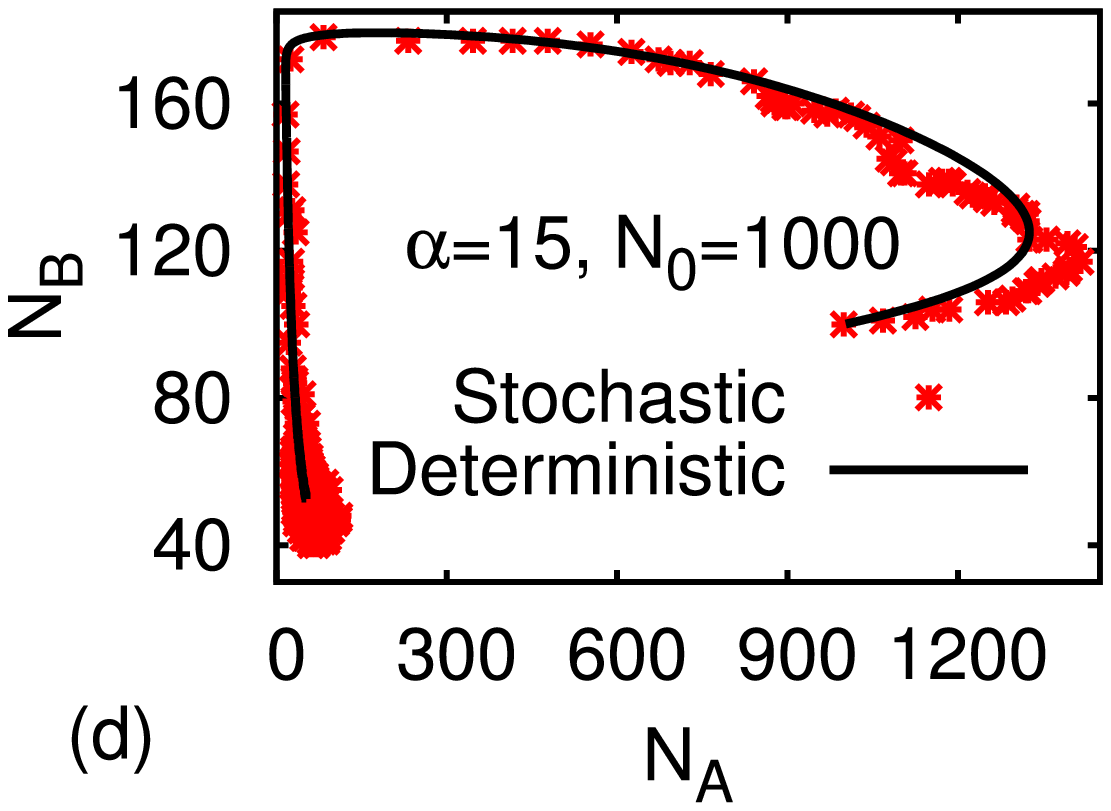}
	\includegraphics*[width=0.33\columnwidth]{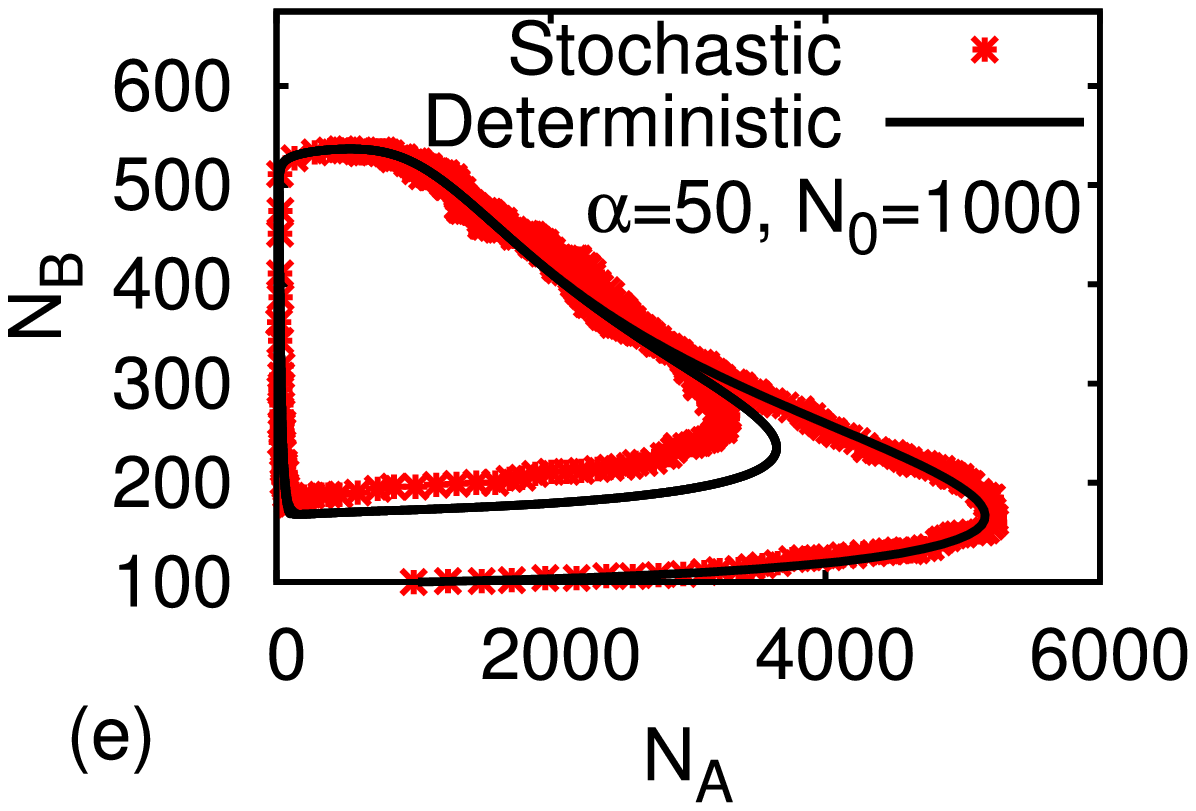}
	\includegraphics*[width=0.341\columnwidth]{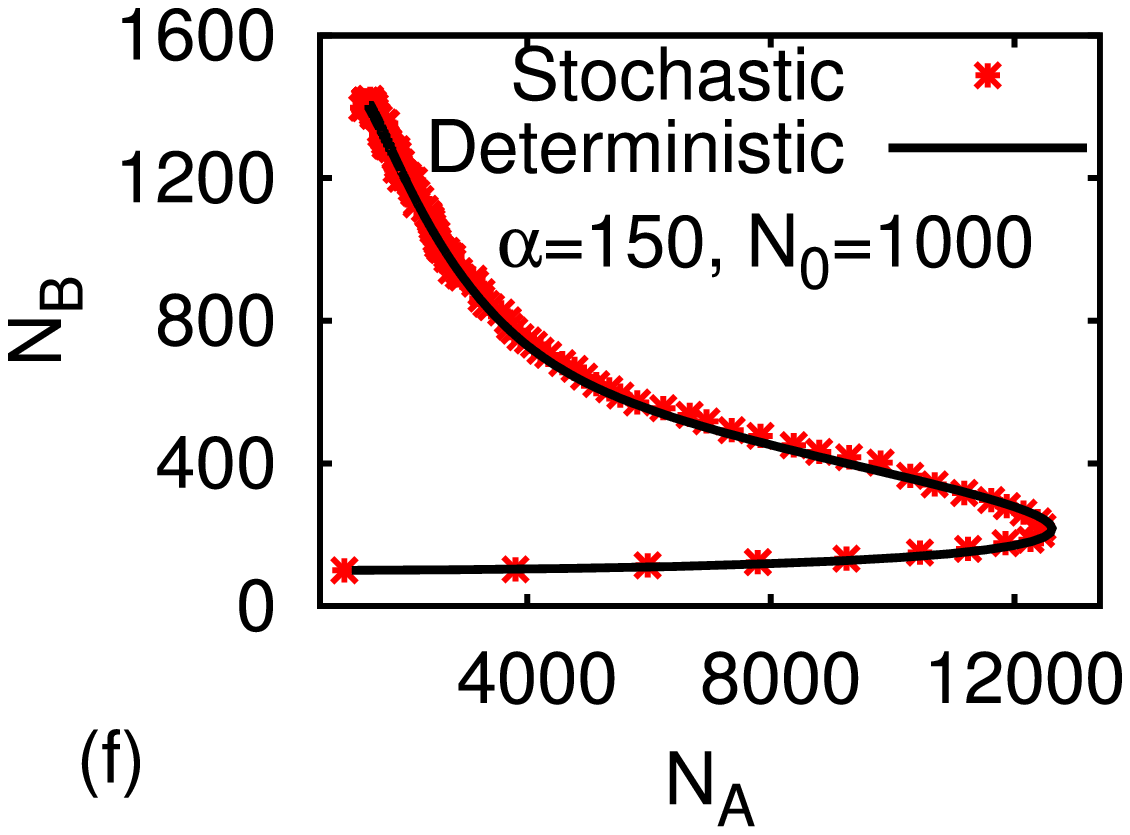}
	\caption{Trajectories in phase space up to $T_G=3000$ (a) , $T_G=10^5$ (b-c) and $T_G=5 \times 10^5$ (d-f) steps of the Gillespie algorithm, respectively, for $N_0=100$ (a-c) and $N_0=1000$ (d-f), in the fixed-point phases $\alpha=15$ (a, d), and $\alpha=150$ (c, f), and in the limit-cycle phase $\alpha=50$ (b, e). The initial condition is $N_A(t=0)=1000$, $N_B(t=0)=100$.}\label{fig2}
\end{figure}

In Fig.~\ref{fig2} we show deterministic (black lines) and stochastic (red symbols) trajectories obtained by numerically solving the deterministic equations (\ref{eq-det11}--\ref{eq-det12}) and via Gillespie simulations, respectively. Similarly to our previous work \cite{hmokaluza}, we use $\alpha$ as the control parameter, with other parameters kept fixed. Typically, we set $K=0.02,b=0.01, \gamma=0.01$ if not stated otherwise. As we vary $\alpha$, we go from the deterministic fixed-point regime for $\alpha<\alpha_1\approx 31.10$ (Fig.~\ref{fig2}a, d), through  the limit-cycle regime (Fig.~\ref{fig2}b, e), to another fixed-point regime for $\alpha>\alpha_2\approx 98.93 $ (Fig.~\ref{fig2}c, f).
\begin{figure}
	\includegraphics*[width=0.5\columnwidth]{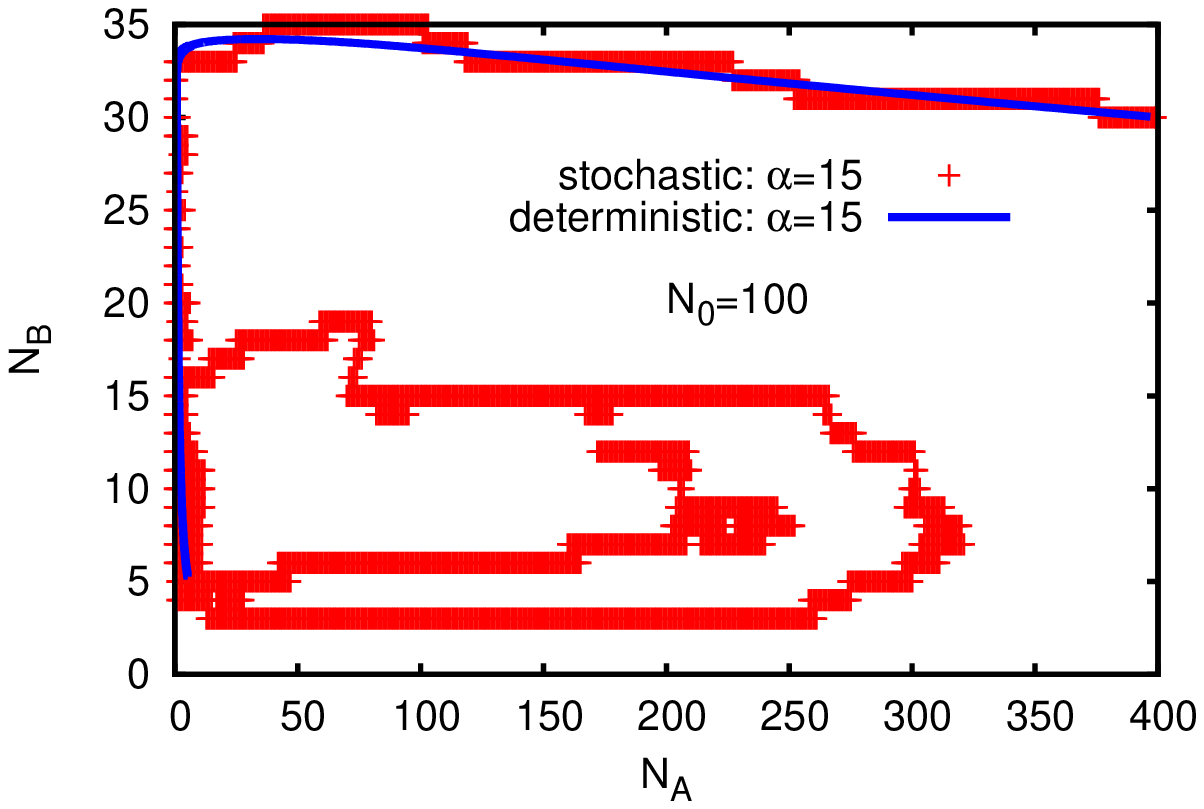}
	\includegraphics*[width=0.5\columnwidth]{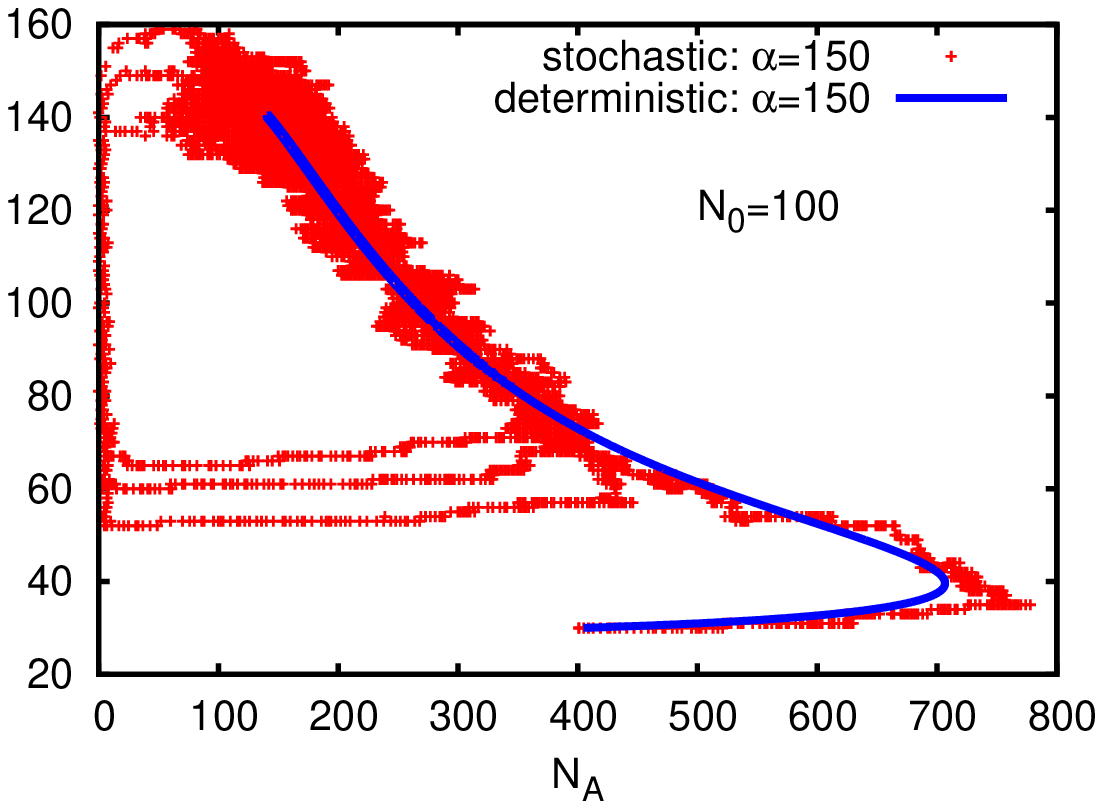} \\
	\center
	\vspace{-1cm}
	\includegraphics*[width=0.5\columnwidth]{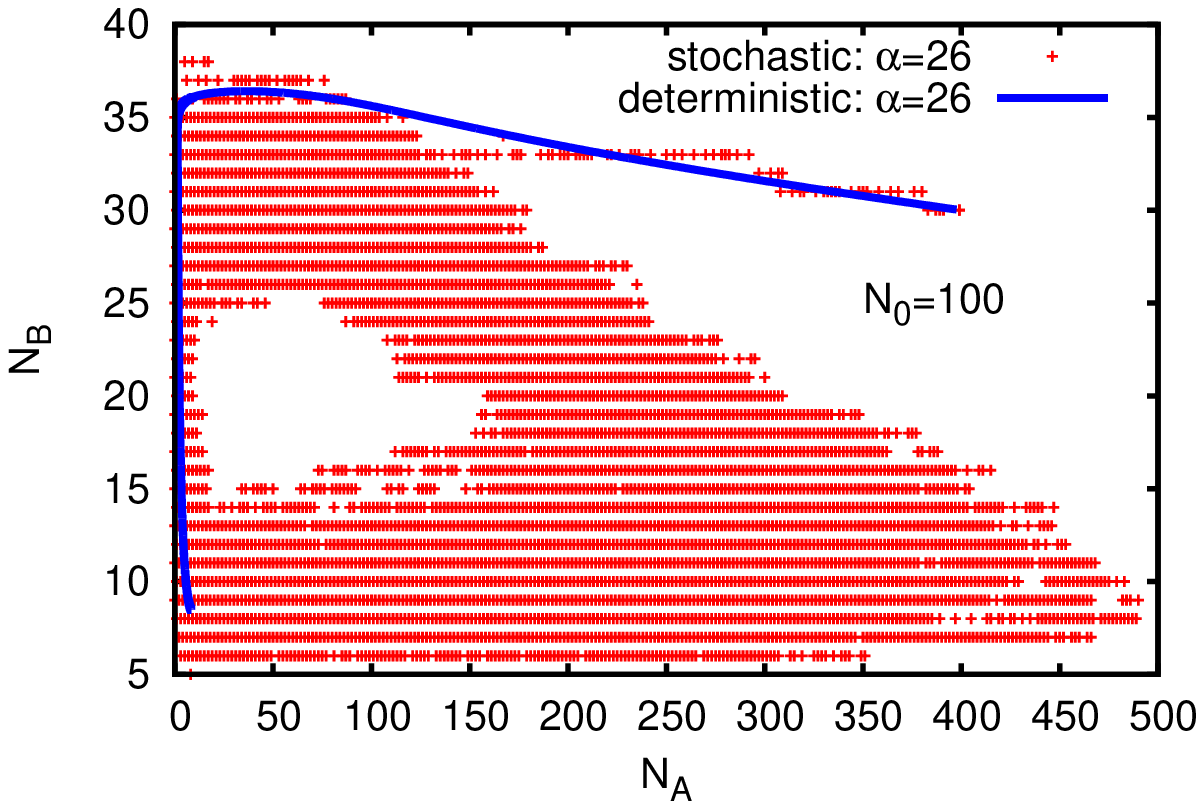}
	\caption{Trajectories in phase space for $T_G=10^4, \alpha=15$ (top left), and $T_G=4\cdot 10^5, \alpha=150$ (top right). Bottom: for $\alpha=26$, close to the transition region, we see cycles for the stochastic simulations (red points) and a trajectory (blue line) converging to the fixed point as deterministic solution. $N_0=100$ in all cases.} \label{fig3}
\end{figure}
In these simulations, after some time the system seems to converge to a neighbourhood of the deterministic fixed point and the deterministic limit cycle, respectively. The stochastic trajectories  roughly follow the deterministic ones, with larger fluctuations for smaller system sizes (as seen from Fig.\ref{fig2}a-c, as compared to Fig.\ref{fig2}d-f). Although it is an abuse to speak of a fixed point and a limit-cycle trajectory in the stochastic system, we shall use these phrases to denote regimes of $\alpha$ which exhibit fixed-point or limit-cycle behaviour for $N_0\to\infty$.

If we, however, wait long enough, we see cycles also deeply in both fixed-point regimes, $\alpha=15$ and $\alpha=150$, see Fig.~\ref{fig3}, top.
As previously mentioned, we shall call such cycles ``quasi-cycles'' due to their absence in the limit $N_0\to\infty$. It should be noticed that $\alpha$ in Fig.\ref{fig3} bottom is close to, but still below the value of the deterministic bifurcation point and accordingly the deterministic trajectory converges to a fixed point, but the stochastic trajectories show large fluctuations and proceed along cycles. The large fluctuations are typical for being in the vicinity of a transition point to another phase. The figure illustrates that it is not possible to visually distinguish the origin of the cycles on the basis of these trajectories. Similarly to finite-size shifts of the critical temperature in phase transitions, one interpretation here would be that the bifurcation point from the fixed-point to the limit-cycle regime  is shifted for finite $N_0$, so that the observed cycles are genuine limit cycles in the limit-cycle regime at finite $N_0$. Alternatively, the stochastic system may be in the fixed-point regime, but with a large number of quasi-cycles. Therefore we need alternative  measures to disentangle the origin of these cycles that we shall consider below.

\begin{figure}
	\centering
		\psfrag{xx}{$\alpha$} \psfrag{yy}{frequency}
		\psfrag{xx2}{$N_0$} \psfrag{yy2}{frequency}
		\includegraphics*[width=12cm]{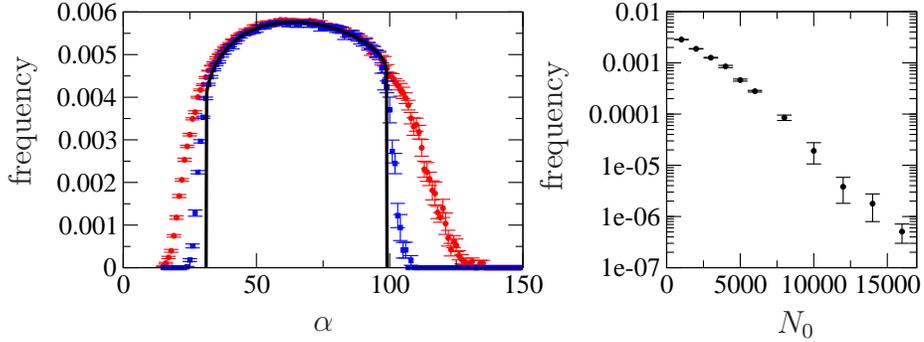}
	\caption{\label{fig_events} Left: Frequency of (quasi-)cycles as a function of $\alpha$. Red circles: $N_0=10^3$, blue squares: $N_0=10^4$. Black line: frequency of oscillations (inverse of the oscillatory period) of the deterministic model. Right: frequency for $\alpha=24$ and different $N_0$. }
\end{figure}

In Fig.~\ref{fig_events}, left, we plot the frequency of such cycles determined in simulations as a function of $\alpha$, for different system sizes $N_0$. The experimentally determined frequency agrees with the frequency of oscillations of the deterministic model in the limit-cycle regime, but it is clearly non-zero for $\alpha$-values which belong to fixed-point regimes\footnote{A similar effect has been observed in Ref.~\cite{hmokaluza} where multiplicative noise is introduced in the deterministic equations (\ref{eq-det11}--\ref{eq-det12}).}. The smaller $N_0$ is, the higher is the frequency of quasi-cycles in these regimes. For fixed $\alpha$, the frequency decays exponentially with increasing $N_0$, which can be seen in Fig.~\ref{fig_events}, right, where we have plotted the frequency for $\alpha=24$ as a function of $N_0$. This suggests that quasi-cycles are large excursions from the fixed point, caused by crossing the inherent ``energy barrier'' which is characteristic for the excitable unit.

\begin{figure}
	\centering
		\psfrag{xx}{$N_A$} \psfrag{yy}{$N_B$}
		\psfrag{gamma001}{$\gamma=0.01$} \psfrag{gamma05}{$\gamma=0.5$}
		\includegraphics*[width=6cm]{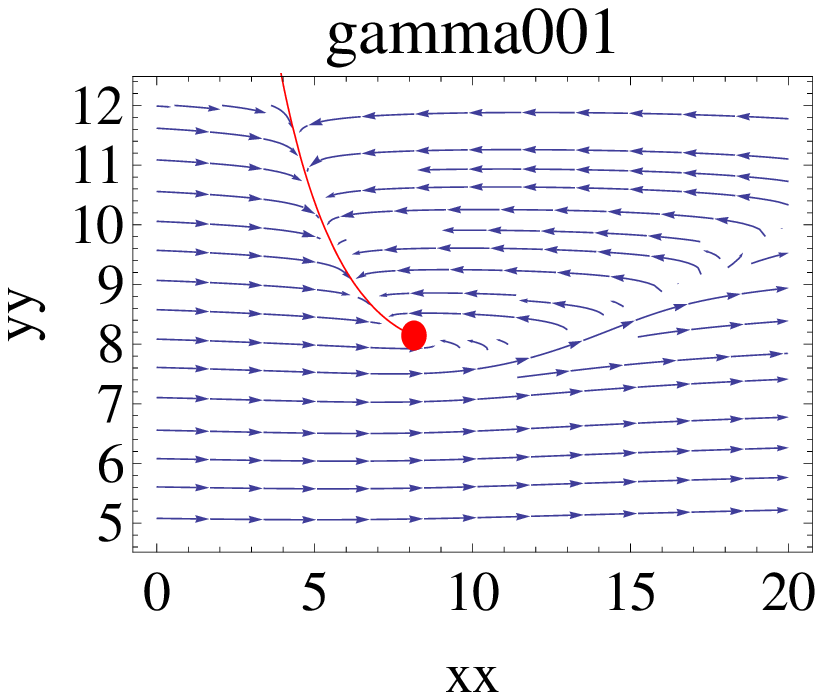}
		\includegraphics*[width=6cm]{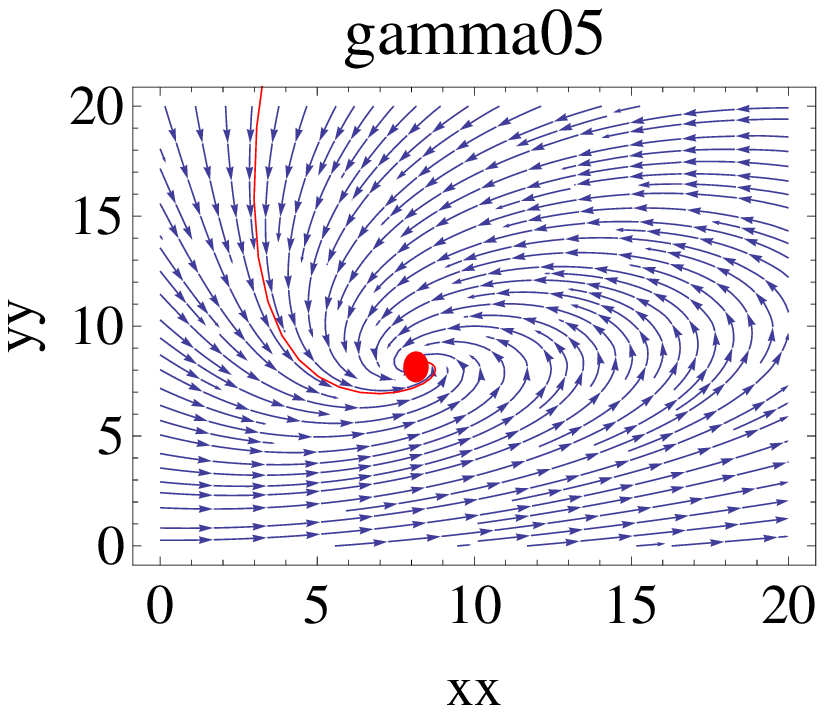}
	\caption{\label{fig_barriers} Plots of the deterministic flow $(dN_A/dt,dN_B/dt)=N_0(d\phi_A/dt,d\phi_B/dt)$ from Eqs.~(\ref{eq-det11}-\ref{eq-det12}) (blue vectors) together with an example of a deterministic solution (red curve) in the fixed-point regime $\alpha=15$. For $\gamma =0.5$ small fluctuations in all directions from the fixed point (end point of the trajectory) lead back to the fixed point after small excursions (oscillations) in phase space, while for $\gamma=0.01$ small fluctuations in almost all directions would directly lead back to the fixed point without any excursion, so that the barrier for  an excursion is very high; only for a small fluctuation in the direction of negative $N_B$ a large excursion would be possible. Correspondingly, the vector field is almost constant over large areas, but changes very rapidly over small areas, in contrast to the flow for $\gamma=0.5$.} \end{figure}

In Fig.~\ref{fig_barriers} we show the deterministic flow $(dN_A/dt,dN_B/dt)$ along with an example of a trajectory which ends in the fixed point. It shall illustrate that the height of the barrier to escape the fixed point increases with decreasing $\gamma$. The plot (left figure) for small $\gamma$ shows that (in almost all directions, apart from that towards decreasing $N_B$), the system can escape the vicinity of the fixed point only if a sufficiently large fluctuation drives it to the regime of small $N_B$/large $N_A$, where the vectors of the flow point outwards from the fixed point. For an excitation beyond such a large barrier the system makes a large excursion in phase space, while it directly relaxes to the fixed point if the excitation is below this threshold. On the other hand, small $N_0$ values help to cross the barrier, because the fluctuations are then larger. The plot for larger $\gamma$ (right figure) illustrates that in this case already small fluctuations are sufficient to escape the fixed point, such fluctuations will then lead to small excursions (cycles) in phase space.

We have also measured the number of quasi-cycles as a function of $\gamma$, other parameters being fixed (results not shown). The increase of the height of the barrier with decreasing $\gamma$ manifests itself in an increasing number of quasi-cycles when $\gamma$ grows.

\begin{figure}
	\centering
		\psfrag{xx}{$t$} \psfrag{yy}{$N_A$}
		\psfrag{xx2}{$\omega$} \psfrag{yy2}{$P_{N_A}(\omega)$}
		\includegraphics*[width=14cm]{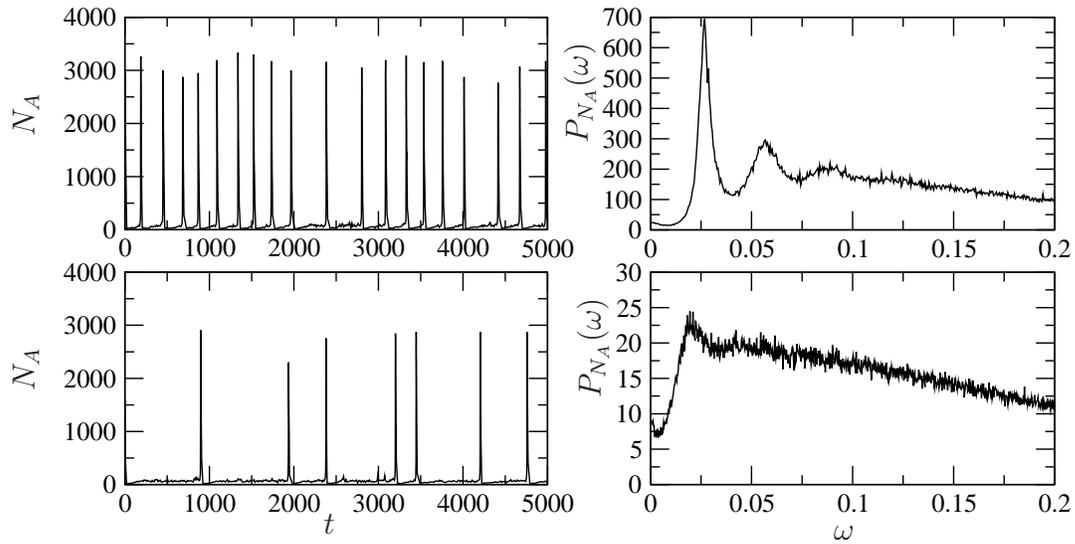}
	\caption{\label{fig_osc}Plots of $N_A(t)$ (left) and their power spectra (right), for $N_0=10^3$ and $\alpha=28$ (top) and $\alpha=20$ (bottom) in the fixed-point regime. Quasi-regular oscillations are visible for $\alpha=28$. For $\alpha=20$, the system still shows spikes in $N_A$, but they are much less regular. }
\end{figure}

We shall show that these cycles, interpreted as quasi-cycles, can occur as quite regular oscillations close to the transition region. In Fig. \ref{fig_osc}, top left, we plot $N_A(t)$ for $\alpha=28$ and $N_0=10^3$. In the same figure, top right, we plot the power spectrum $P_{N_A}(\omega) = \left|\int N_A(t) e^{-i t \omega} dt\right|^2$ which shows a peak at $\omega=0.027$. This clearly indicates the regular, oscillatory nature of quasi-cycles for $\alpha$ close to the transition point. As $\alpha$ gets smaller, these oscillations become less regular and finally become rare stochastic events, as shown in the same figure, bottom.

\begin{figure}
\includegraphics*[width=0.5\columnwidth]{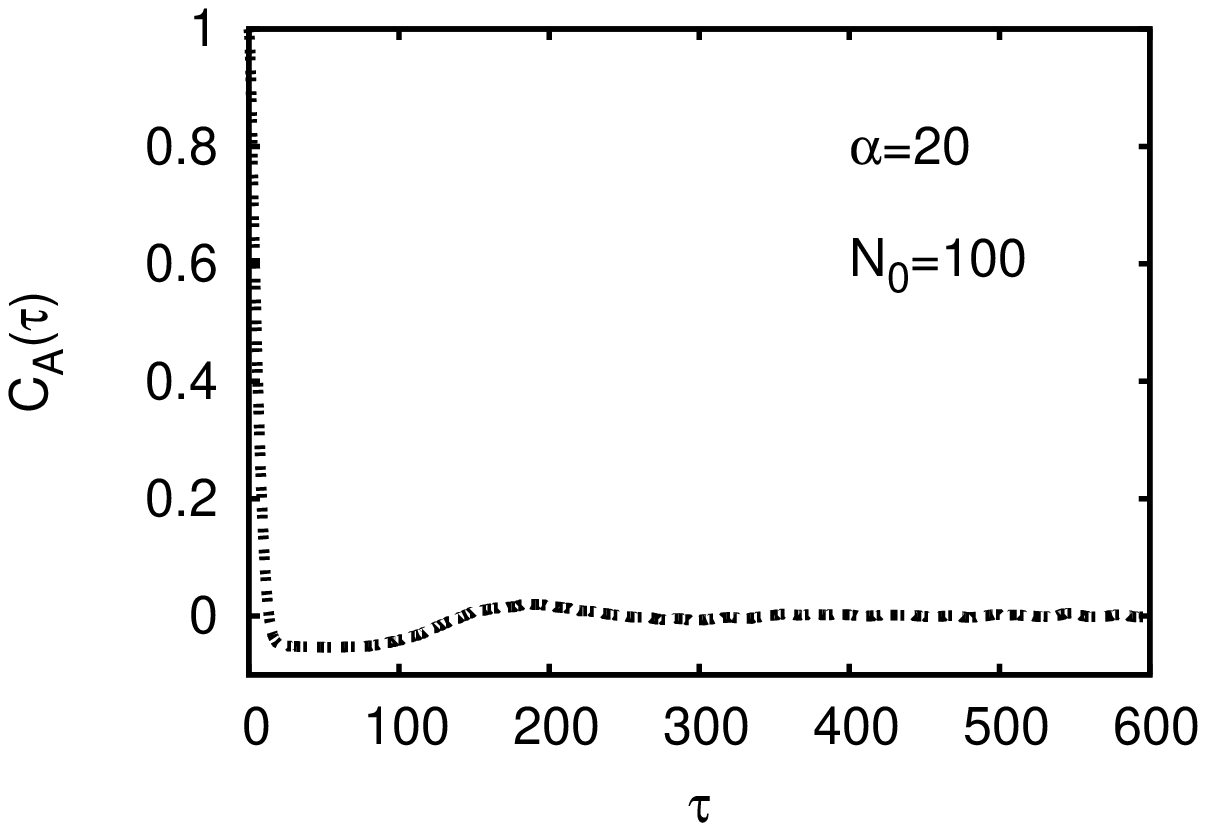}
\includegraphics*[width=0.5\columnwidth]{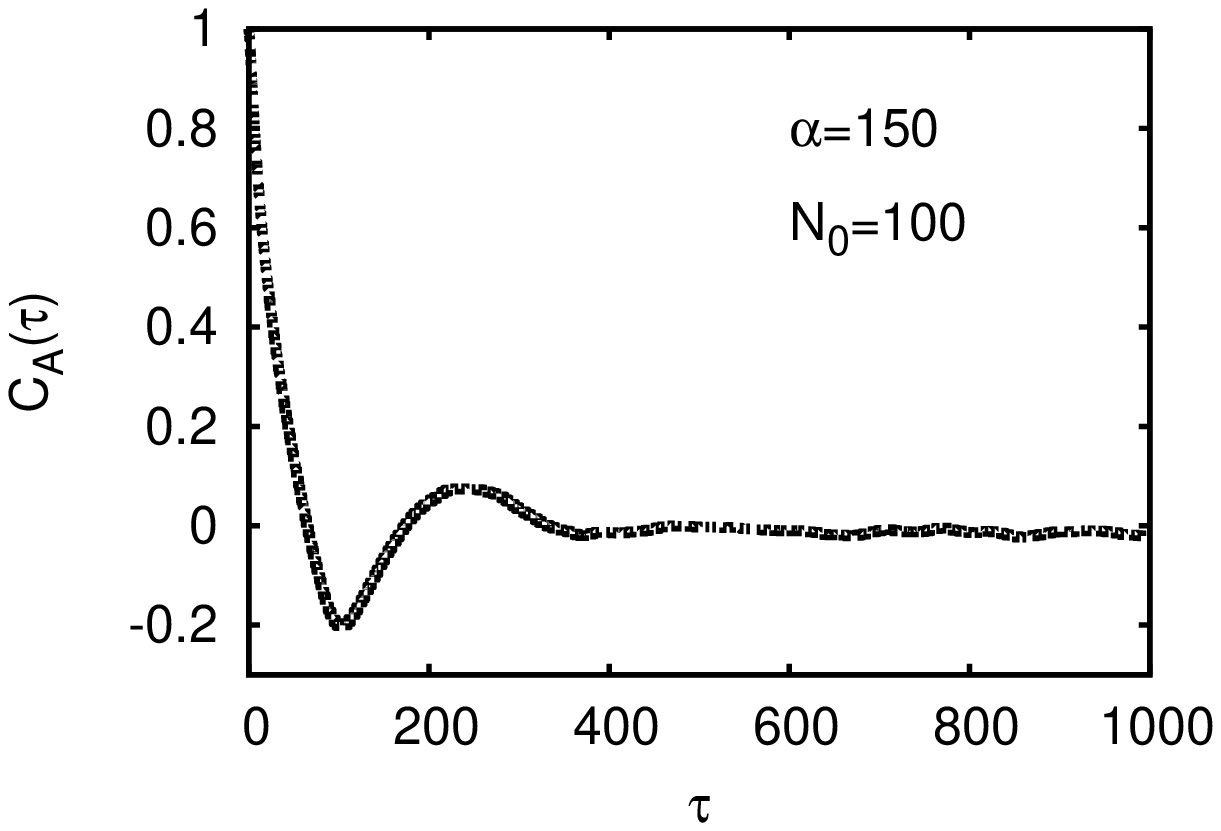}
\center
\includegraphics*[width=0.5\columnwidth]{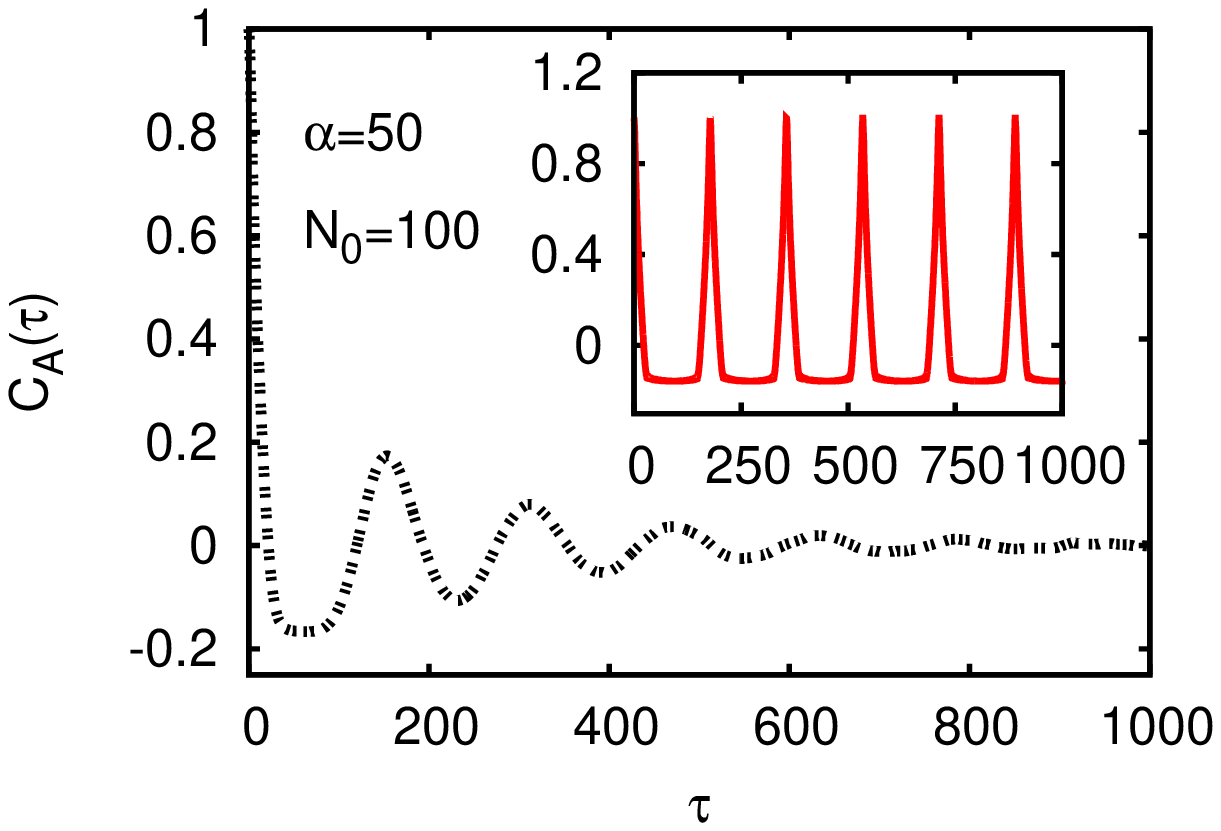}
\caption{\label{fig5}Autocorrelation functions $C_A(\tau)$ for $N_A(t)$ for different values of $\alpha=20,50,150$ and $N_0=100$. The inset for $\alpha=50$ shows the deterministic behaviour of $C_A(\tau)$ in the limit-cycle regime. The maxima for $\tau>0$ indicate the presence of (quasi-)cycles.}
\end{figure}
\begin{figure}
\center
\includegraphics*[width=0.7\columnwidth]{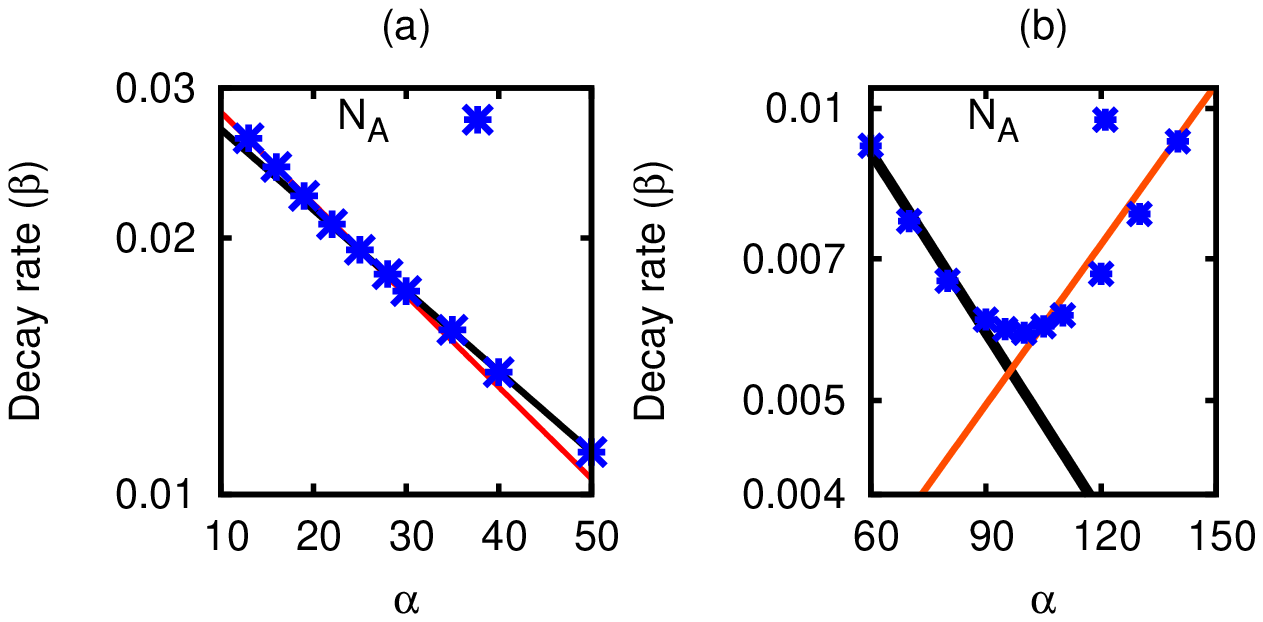}
\caption{Decay rate $\beta$ of the autocorrelation function of $N_A$ as function of $\alpha$, for $N_0=100, \gamma=0.01$. The rate $\beta(\alpha)$ changes its slope at $\alpha=26\pm 6$ and $\alpha=96\pm 9$
which agrees with the lower and the upper bifurcation point 
$\alpha_1\approx 31.10,\alpha_2\approx 98.93$.} \label{fig6}
\end{figure}

In the following we shall show how we can distinguish genuine limit cycles from quasi-cycles by comparing decay rates of autocorrelation functions of $N_A(t),N_B(t)$ obtained from simulations. For simplicity, we shall focus only on $N_A$. In Fig.~\ref{fig5} we plot the autocorrelation function:
\begin{equation}
 C_A(\tau) = \frac{\int_0^{t_{\rm max}} (N_A(t)-\langle N_A\rangle)(N_A(t+\tau)-\langle N_A\rangle) dt}{\int_0^{t_{\rm max}} (N_A(t)-\langle N_A\rangle)^2 dt}. \label{acfd1}
\end{equation}
Here $t_{\rm max}$ denotes the length of the sample in time (optimally, as long as possible), and the value of $N_A(t)$ obtained from the Gillespie simulation is such that $N_A(t)=N_A(t_i)$ for the largest $t_i$ such that $t_i\leq t$. We stress that $C_A(\tau)$ measures autocorrelations in the real, physical time and not in the number of steps of the Gillespie algorithm.
In practice, we have calculated $C_A(\tau)$ from the numerically obtained time series $\{N_A(t_i),t_i\}$ by resampling them at uniform time intervals, performing the Fast Fourier Transform (FFT), squaring the absolute values of the coefficients, and doing the inverse FFT. This method is not only much faster than calculating $C_A(\tau)$ directly from Eq.~(\ref{acfd1}), but it also produces the power spectrum $P(\omega)$ of the signal as a by-product, which we also use later in this work.
Figure~\ref{fig5} shows examples of $C_A(\tau)$ for all three regimes of $\alpha$. One sees a number of maxima for $\tau>0$ which indicate the presence of cycles and quasi-cycles. The amplitudes of these peaks decay exponentially with $\tau$. If we fit the exponential decay $e^{-\beta \tau}$ to these peaks and plot the decay rate $\beta$ as function of $\alpha$ as in Fig.~\ref{fig6}, we see that $\beta$ changes faster with $\alpha$ for quasi-cycles than for limit cycles. This behaviour seems to be a generic feature of oscillating stochastic systems \cite{golden2,bratsun}. If we now fit straight lines to $\beta(\alpha)$ in different regimes (see Fig.~\ref{fig6}), we see that these lines cross very close to the deterministic transition points $\alpha_1,\alpha_2$.

\begin{figure}
\psfrag{xx}{$t$} \psfrag{yy}{$N_A(t)$}
\includegraphics*[width=\columnwidth]{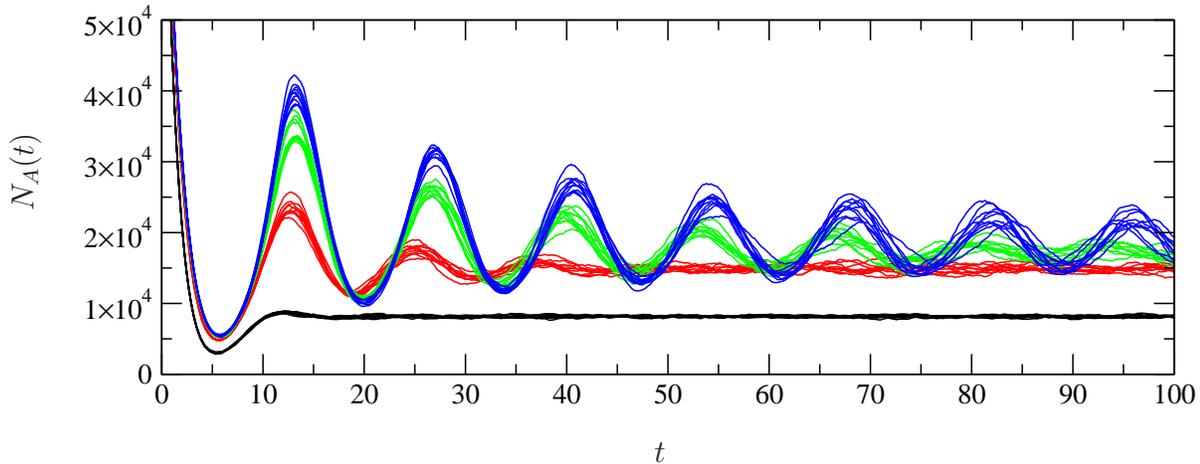}
\caption{\label{fig7}Plots of $N_A(t)$ for $N_0=10^5,\gamma=0.5$ and different $\alpha$ from the fixed-point regime: 25 (black), 40 (red) 43 (green) and 44 (blue), curves from bottom to top. For each $\alpha$, 10 independent realisations of $N_A(t)$ starting from the same initial condition are shown. As $\alpha$ approaches the transition point $\alpha\approx 45$ to the limit-cycle regime, the coherence time between independent realisations increases. This indicates that relaxation to the stationary state slows down at the transition point.}
\end{figure}

We shall now show that these transition points, which were identified as subcritical Hopf bifurcations in the deterministic limit, have precursors in the stochastic model. In general, such precursors are larger fluctuations, coherent behaviour in space and/or time, and phenomena similar to critical slowing down known from second order phase transitions. In Fig. \ref{fig7}, we plot sets of 10 independent trajectories of $N_A(t)$ for $N_0=10^5$, $\gamma=0.5$, starting from the same initial state, each set for another value of $\alpha$. We observe an increased coherence in time: as $\alpha$ approaches the transition at $\alpha_1(\gamma=0.5)\approx 45$ from the fixed-point to the limit-cycle regime, independent time series follow each other for an increasing period of time. For example, the time series for $\alpha=44$ follow the very same oscillating trajectory up to $T > 100$, while away from the transition region ($\alpha=25$),  trajectories reach the stationary state already after $T\approx 20$.
To observe the coherence, trajectories have to stay close to the deterministic solution. This is fulfilled if the number of particles is sufficiently large. For $\gamma=0.5$, it is sufficient to choose $N_0=10^5$. For our usual parameter choice of $\gamma=0.01$, it is necessary to take $N_0\geq 10^6$ to see the coherence, and even then the increase in the coherence time is seen only in a very narrow region below the critical value $\alpha_1\approx 31.1$ (results not shown).

\begin{figure}
\includegraphics*[width=0.5\columnwidth]{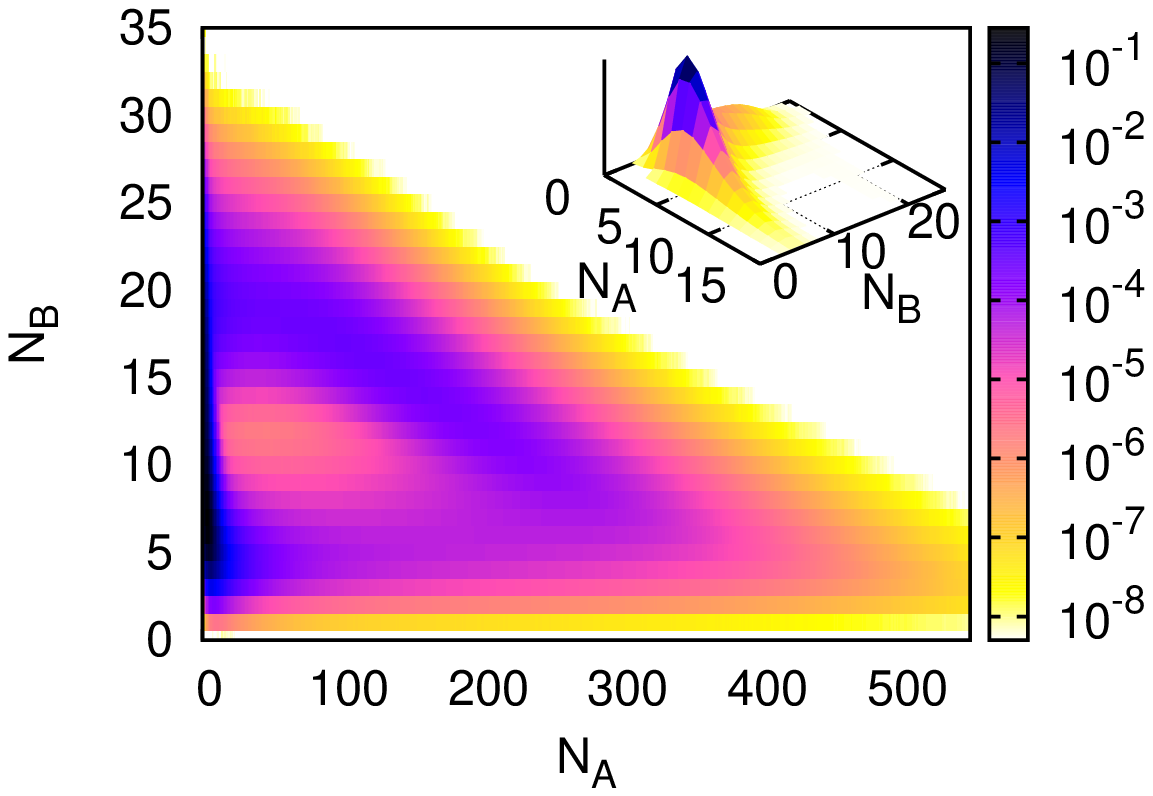}
\includegraphics*[width=0.5\columnwidth]{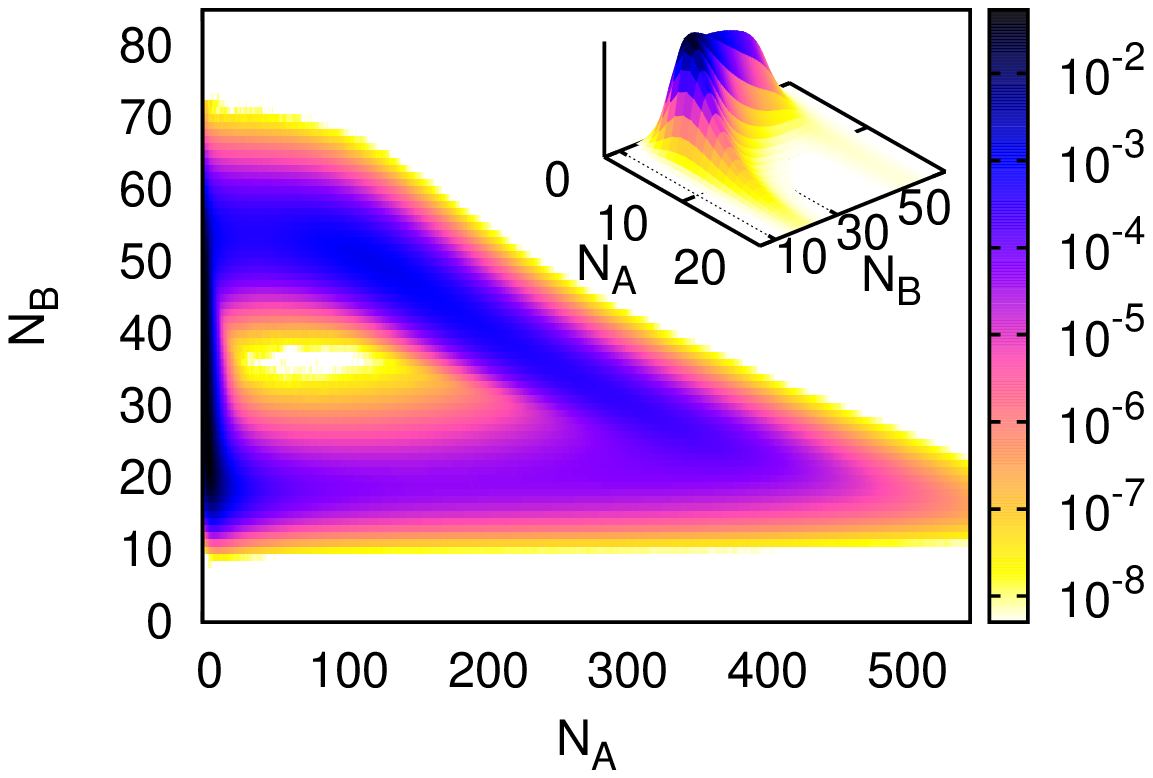}
\caption{Density plots of the probability $P^*(N_A,N_B)$ in the stationary state, for $\alpha=15$ (fixed-point regime, left) and $\alpha=50$ (limit cycle, right), for $N_0=100$. Insets show peaks in the probability from small-$N_A$ regions.} \label{fig8}
\end{figure}

In the last part of this section, we shall discuss the stationary probability distribution $P^*(N_A,N_B)=P(N_A,N_B,t\to\infty)$ of finding the particle at $(N_A,N_B)$ after the system has forgotten the initial condition. The plot is shown in Fig.~\ref{fig8} for two values of $\alpha$, in the fixed-point and in the limit-cycle regime. The support of the probability distribution has always a donut-like shape and a maximum for small values of $N_A$. The relative amplitudes of the donut and the peak depend on $\alpha$. The peak corresponds to the probability localized around the fixed point, whereas the donut shape corresponds to cycles and quasi-cycles.
We also see from Fig.~\ref{fig8} that the range of fluctuations strongly depends on the location along the limit cycle: it is very large for large values of $N_A$ and small for small values of $N_A$. In the next section we shall analytically predict how the variance of $N_A$ evolves in time, and how the autocorrelations decay in time for different regimes of $\alpha$.

\section{Analytical approach}
\label{sec4}

So far, we have shown the results of computer simulations.
We cannot find analytical solutions of the master equation in a closed form, nor is a product ansatz $P(N_A,N_B)=P(N_A)P(N_B)$ justified, since both variables are not on an equal footing with respect to their coupling in our model. However, we shall see that the van Kampen expansion \cite{vankampen} in the inverse of the system size $N_0$ captures many (but not all) of the observed features. This method assumes that we can split the integer variables $N_A,N_B$ into a deterministic part proportional to $N_0$, and a stochastic part proportional to $\sqrt{N_0}$:
\begin{eqnarray}
N_A(t) &=& N_0\phi_A(t)\;+\;\sqrt{N_0}\xi(t) \label{eqtrafo1} \\
N_B(t) &=& N_0\phi_B(t)\;+\;\sqrt{N_0}\eta(t) \label{eqtrafo2},
\end{eqnarray}
where fluctuations $\xi$ of $N_A$ and $\eta$ of $N_B$ are continuous variables of order $O(1)$. We now introduce the step operators
\begin{equation}
\mathbb{E}_{n}^+[g(n)]= g(n+1), \qquad
\mathbb{E}_{n}^-[g(n)]= g(n-1), \label{eq-stop}
\end{equation}
acting on an arbitrary function $g(n)$. The master equation (\ref{eq-master}) can then be rewritten as follows:
\begin{eqnarray}
 \frac{\partial P(N_A, N_B)}{\partial t}&=& (\mathbb{E}^-_{N_A}-1)N_0f(N_A/N_0,N_B/N_0)P(N_A,N_B) \nonumber \\
 &+&(\mathbb{E}^+_{N_A}+\gamma \mathbb{E}^{-}_{N_B}-\gamma -1)N_AP(N_A,N_B) \nonumber\\
 &+& (\mathbb{E}^+_{N_B}-1)\gamma N_B P(N_A,N_B),
\label{eq-masterop}
\end{eqnarray}
where $f(N_A/N_0,N_B/N_0)$ is defined in Eq.~(\ref{eq-f}).
For given, yet undetermined functions $\phi_A(t),\phi_B(t)$, Eqs. (\ref{eqtrafo1}--\ref{eqtrafo2}) correspond to a transformation of variables $(N_A,N_B)$ to $(\xi,\eta)$. Therefore the function $P(N_A,N_B,t)$ defines a new function $\Pi(\xi,\eta,t)$. This way the time derivative of $P(N_A,N_B,t)$ can be written in terms of the new variables as
\begin{equation}
 \frac{\partial P}{\partial t}=\frac{\partial \Pi}{\partial t}- \sqrt{N_0}\frac{d\phi_A}{dt}\frac{\partial \Pi}{\partial \xi}-\sqrt{N_0}\frac{d\phi_B}{dt}\frac{\partial \Pi}{\partial \eta}.
\label{eq-p2pi}
\end{equation}
If we make a Taylor expansion of the step operators in powers of $N_0^{-1/2}$,
\begin{eqnarray}
 \mathbb{E}^\pm_{N_A}&=&1\pm N_0^{-\frac{1}{2}}\frac{\partial}{\partial \xi}+\frac{1}{2}N_0^{-1}\frac{\partial^2}{\partial \xi^2}+... \\
 \mathbb{E}^\pm_{N_B}&=&1\pm N_0^{-\frac{1}{2}}\frac{\partial}{\partial \eta}+\frac{1}{2}N_0^{-1}\frac{\partial^2}{\partial \eta^2}+...\;,
\end{eqnarray}
and insert these expressions into the master equation (\ref{eq-masterop}), we obtain to order $N_0^{1/2}$ the deterministic equations (\ref{eq-det11}--\ref{eq-det12}) and to order $N_0^0=1$ the following Fokker-Planck equation:
\begin{equation}
 \frac{\partial \Pi ({\bf x},t)}{\partial t} = -\sum_{i,j\in\{1,2\}}A_{ij}\frac{\partial}{\partial x_i}x_j\Pi({\bf x},t) +\frac{1}{2}\sum_{i,j}B_{ij}\frac{\partial \Pi({\bf x},t)}{\partial x_i\partial x_j},
\label{eq-pi-b}
\end{equation}
where ${\bf x}=(x_1,x_2)=(\xi,\eta)$, and $A_{ij}$ and $B_{ij}$ are the following matrices
\begin{equation}\label{eqA}
 A = \left[\begin{array}{cc} -1+f_A & f_B \\ \gamma & -\gamma \end{array}\right],
\end{equation}
and
\begin{equation}\label{eqB}
 B = \left[\begin{array}{cc} \phi_A + f & 0 \\ 0 & \gamma(\phi_A + \phi_B) \end{array}\right].
\end{equation}
Here we have introduced the notation $f\equiv f(\phi_A(t),\phi_B(t))$, $f_A \equiv \frac{\partial f(\phi_A(t),\phi_B(t))}{\partial \phi_A(t)}$, and $f_B \equiv \frac{\partial f(\phi_A(t),\phi_B(t))}{\partial \phi_B(t)}$.
This Fokker-Planck equation is known to have a Gaussian distribution as unique solution \cite{vankampen}, so it is sufficient to determine its first and second moments. From Eq.~(\ref{eq-pi-b}) we derive the following equations for the first moments
\begin{eqnarray}
 \frac{d\langle\xi\rangle}{dt} &=& \left(f_A-1\right)\langle\xi\rangle+f_B\langle\eta\rangle,  \label{eqfirstmoments1} \\
 \frac{d\langle\eta\rangle}{dt} &=& \gamma (\langle\xi\rangle - \langle\eta\rangle), \label{eqfirstmoments2}
\end{eqnarray}
with $\langle\xi\rangle = \int \xi \Pi(\xi, \eta,t) d\xi d\eta$ and $\langle\eta\rangle = \int \eta \Pi(\xi, \eta,t) d\xi d\eta$. For the second moments we obtain
\ba
\label{eq2mom}
\frac{d}{dt}\left[ {\begin{array}{c}
                     \langle\xi^2\rangle \\
		     \langle\eta^2\rangle \\
		    \langle\xi \eta\rangle
                    \end{array}} \right]
 &=&
\left[ {\begin{array}{ccc}
2(f_A-1) & \quad 0 & \quad\quad 2f_B \\
0 & \quad -2 \gamma & \quad\quad 2\gamma \\
\gamma & \quad f_B & \quad\quad (f_A - 1 - \gamma)
\end{array} } \right]\left[ {\begin{array}{c}
                     \langle\xi^2\rangle \\
		     \langle\eta^2\rangle \\
		    \langle\xi \eta\rangle
                    \end{array}} \right] \nonumber \\
                    &+& \left[ {\begin{array}{c}
                     \phi_A+f \\
		     \gamma(\phi_A+\phi_B) \\
		     0
                    \end{array}} \right].
\ea
We shall show now that these equations correctly describe fluctuations not only in the fixed-point regime, but also in the limit-cycle regime, provided that $N_0$ is sufficiently large and we are not too close to the transition (bifurcation) points. We shall also show how this approximation breaks down in the vicinity of transition points, and explain the reason why the van Kampen expansion does not work in these cases. Therefore, we will investigate how the variances of $N_A,N_B$ evolve in time, calculate the autocorrelation function and the power spectrum in the stationary state in both regimes, and compare these quantities with Gillespie simulations.

\subsection{Solutions in the fixed-point phase}
\begin{figure}
	\centering
	\psfrag{xx}{$\alpha$}\psfrag{yy}{$\left<\xi^2\right>_s$}
		\includegraphics*[width=14cm]{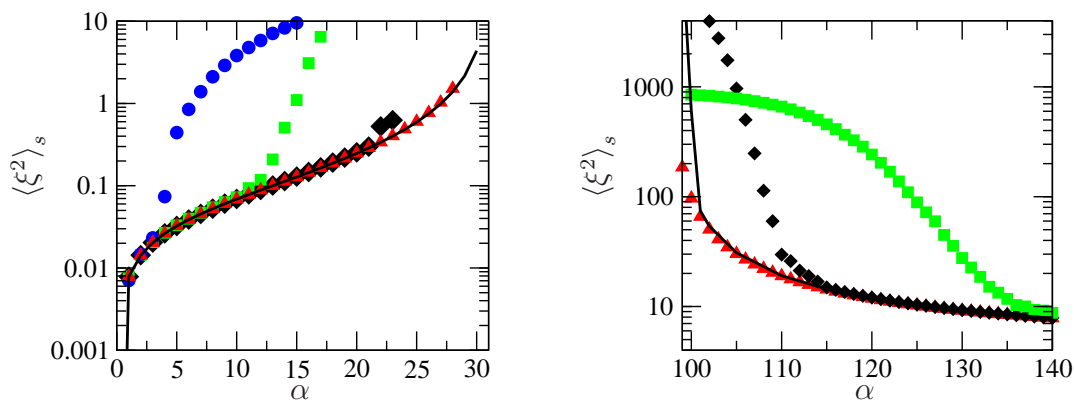}
	\caption{\label{fig_xi2_fixed}Stationary-state variance $\left<\xi^2\right>_s$ as a function of $\alpha$, obtained from Eq.~(\ref{eq-xisqs1}) (solid line) and from Gillespie simulations for $N_0=100$ (blue circles), $10^3$ (green squares), $10^4$ (black diamonds) and $10^5$ (red triangles). }
\end{figure}

Let us first discuss the solution for $\alpha$ from the fixed-point phase. Assuming that the system starts always from the same point in the phase space so that the averages $\langle\xi(0) \rangle=0$ and $\langle\eta(0) \rangle=0$ are zero at time $t=0$, the linearity of Eqs.~(\ref{eqfirstmoments1}--\ref{eqfirstmoments2}) causes the averages to stay equal to zero for any time $t>0$.
Therefore, the equations for the second moments are actually the equations for the variances. In the stationary state, the time derivatives on the left-hand side of Eq.~(\ref{eq2mom}) vanish, and $\phi_A=\phi_B\equiv \phi^*$, where $\phi^*$ can be determined numerically from Eq.~(\ref{eq-det11}) as the root of $\phi^*=f(\phi^*,\phi^*)$. We then obtain from Eq.~(\ref{eq2mom}) that the stationary second moments of the fluctuations are
 \begin{eqnarray}
  \langle\xi^2\rangle_s &=& 2\phi^*\frac{ \gamma+1-f_A - f_B +  f_B^2}{D}, \label{eq-xisqs1}\\
 \langle\eta^2\rangle_s &=& \phi^*+4\phi^* \frac{\gamma + (1-f_A)f_B}{D}, \label{eq-xisqs2}\\
 \langle\xi\eta\rangle_s&=& 2\phi^* \frac{\gamma+(1-f_A)f_B}{D}, \label{eq-xisqs3}
\end{eqnarray}
where
\bq
	D = 2(\gamma+1-f_A)(1- f_A -  f_B).
\eq
The subscript ``$s$'' stands for the stationary state. In this set of equations, the quantities $\phi_A$, $\phi_B$, $f_A=\frac{df}{d\phi_A}$ and $f_B=\frac{df}{d\phi_B}$ should be evaluated at the fixed point which depends on the choice of parameters $\gamma,b,K,\alpha$. In Fig. \ref{fig_xi2_fixed} we compare the variance (\ref{eq-xisqs1}) with $\langle\xi^2\rangle_s = (\langle N_A^2\rangle - \langle N_A \rangle^2)/ N_0$ obtained from the Gillespie simulations. The agreement is very good if $\alpha$ is deep in the fixed-point regime, but it becomes worse as $\alpha$ approaches any of the transition points. This is caused by the more and more frequent occurrence of large quasi-cycles, induced by large fluctuations, not captured by the above approximation, which increase the variance by orders of magnitude.

\begin{figure}
\centering
\psfrag{xx}{$\tau$} \psfrag{yy}{$C_A(\tau)$}
\includegraphics*[width=14cm]{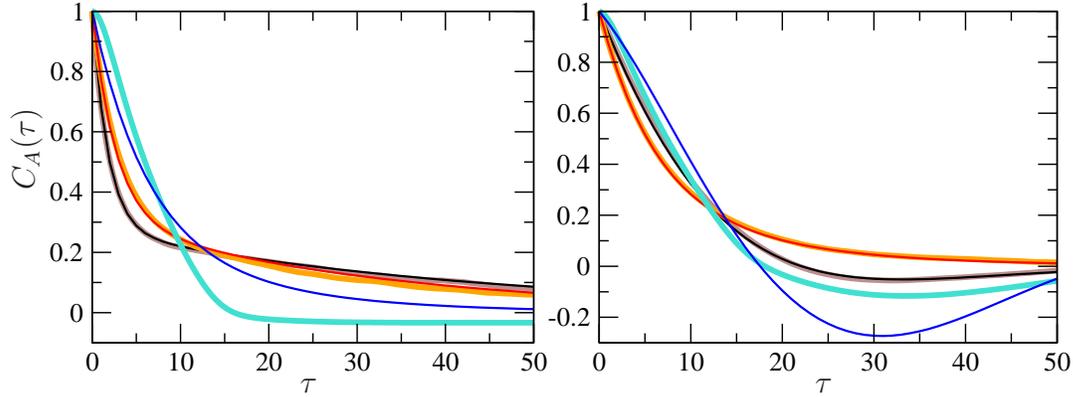}
\caption{\label{fig9}Comparison between autocorrelation function $C_A(\tau)$ from Eq.~(\ref{corrtheor}) (thin lines) and the results of Gillespie simulations (thick lines). Left: $N_0=5000$, $\alpha=15,20,25$ (black/grey, red/orange, blue/turquoise). Right: $N_0=10^5$, $\alpha=25,27,29$ (black/grey, red/orange, blue/turquoise). For the same value of $\alpha=25$, the agreement gets better when $N_0$ increases.}
\end{figure}

In a second step we would like to derive the autocorrelation function for $\xi(t)$ from the solutions of the moment equations. The differential equations for the time evolution of autocorrelation functions $\langle \xi(0)\xi(\tau)\rangle$ and $\langle \xi(0)\eta(\tau)\rangle$ are the same as for the averages (\ref{eqfirstmoments1}--\ref{eqfirstmoments2}), since $\xi(0)$ is constant from a $\tau$'s point of view, but the initial condition must take into account fluctuations around the fixed point, which do not average out to zero when combined with $\xi(\tau)$. We therefore have:
\begin{eqnarray}
 \frac{d\langle\xi(0)\xi(\tau)\rangle}{d\tau} &=& \left(f_A-1\right)\langle\xi(0)\xi(\tau)\rangle+f_B\langle\xi(0)\eta(\tau)\rangle, \\
 \frac{d\langle\xi(0)\eta(\tau)\rangle}{d\tau} &=& \gamma (\langle\xi(0)\xi(\tau)\rangle - \langle\xi(0)\eta(\tau)\rangle),
\end{eqnarray}
with $\langle\xi(0)\xi(0)\rangle = \langle\xi^2\rangle_s, \langle\xi(0)\eta(0)\rangle = \langle\xi\eta\rangle_s$ from Eqs.~(\ref{eq-xisqs1}), (\ref{eq-xisqs3}) as the initial condition.
In the following we focus only on the autocorrelations of $\xi(t)$. As $\tau$-dependent solutions we obtain
\ba
 \langle\xi(0)\xi(\tau)\rangle &=& \frac{\langle\xi^2\rangle_s}{2} \left(e^{\tau\lambda_1} + e^{\tau\lambda_2}\right) \nonumber \\
 &+& \frac{2\langle\xi\eta\rangle_s f_B + \langle\xi^2\rangle_s (\gamma-1+f_A)}{2(\lambda_1-\lambda_2)} \left(e^{\tau\lambda_1} - e^{\tau\lambda_2}\right), \label{eq-etat}
\ea
where $\lambda_{1,2}$ are given by
\bq
\lambda_{1,2}=-\frac{\gamma-f_A+1}{2} \pm \frac{1}{2}\sqrt{(\gamma-f_A+1)^2+4\gamma(f_A+f_B-1)}.
\label{eq-lambda}
\eq
Again, $\phi_A,\phi_B$-dependent quantities should be evaluated at the fixed point. In Fig.~\ref{fig9} we compare the numerically obtained $C_A(\tau)$ from Eq.~(\ref{acfd1}) with
\bq
	C_A(\tau) = \frac{\langle N_A(0)N_A(\tau)\rangle-\langle N_A(0)\rangle^2}{\langle N_A^2(0)\rangle-\langle N_A(0)\rangle^2} = \frac{\langle\xi(0)\xi(\tau)\rangle}{\langle\xi^2(0)\rangle}, \label{corrtheor}
\eq
where $\langle\xi(0)\xi(\tau)\rangle$ is calculated from Eq.~(\ref{eq-etat}). Similarly to what we have seen for the variance, the range of $\alpha$ over which the above formula applies increases with $N_0$.

\begin{figure}
\centering
\psfrag{xx}{$\omega$} \psfrag{yy}{$P_\xi(\omega)$}
\includegraphics*[width=0.85\columnwidth]{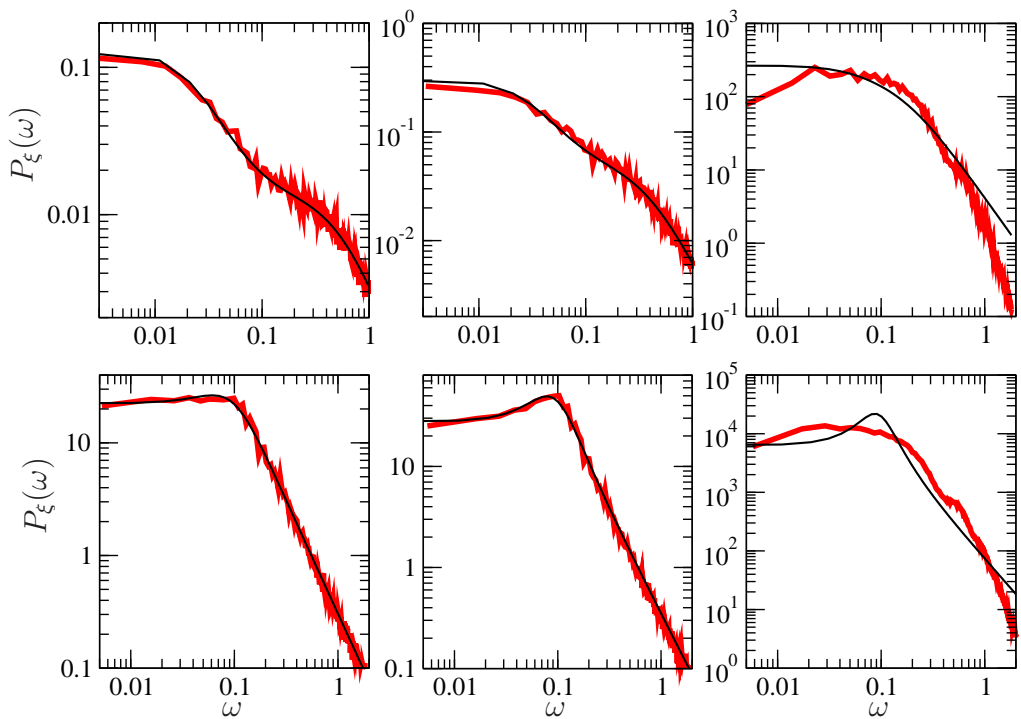}
\caption{ \label{fig10}Comparison between the power spectrum $P_\xi(\omega)$ in the fixed point from Eq.~(\ref{eqpo1}) and obtained in simulations for $N_0=5000$ (top) and $N_0=10^5$ (bottom). Top left to right: $\alpha=15,20,25$. Bottom left to right: $\alpha=27,28,29$.}
\end{figure}

The power spectrum of the fluctuations can be either obtained via the Fourier transform of the corresponding autocorrelations, or by directly Fourier-transforming the Langevin equation for $\xi(t), \eta(t)$ that corresponds to the Fokker-Planck equation (\ref{eq-pi-b}) \cite{vankampen}. In our case, the Langevin equation is given by
\begin{equation}
 \frac{d{\bf x}(t)}{dt} = A^{*}{\bf x}(t)+{\bf \zeta}(t),
\label{eq-lan}
\end{equation}
where $\bf x=\{\xi,\eta\}$, the drift matrix $A^{*}$ from Eq.~(\ref{eqA}) is evaluated at the fixed point $\phi_A=\phi_B=\phi^*$, and $\zeta(t) = \{\zeta_1(t), \zeta_2(t)\}$ is a bivariate white Gaussian noise with
\begin{eqnarray}
 \Big<\zeta_i(t)\Big> &=&0, \\
 \Big<\zeta_i(t)\zeta_j(t^{'})\Big> &=& B^{*}_{ij}\delta(t-t^{'}),
\end{eqnarray}
where $i,j\in\{1,2\}$ and the matrix $B^*$ is given by Eq.~(\ref{eqB}) evaluated at the fixed point. As before, we shall focus on $\xi(t)$. We transform Eq.~(\ref{eq-lan}) to the Fourier space using $\tilde{x_i}(\omega) = \int^{+\infty}_{-\infty}x_i(t)e^{-i\omega t} dt$, solve the resulting algebraic equations for $\tilde{\xi},\tilde{\eta}$, and calculate the power spectrum $P_\xi(\omega) = \langle|\tilde{\xi}(\omega)|^2\rangle$, where the average can be expressed in terms of the coefficients $B^*_{ij}$. Remembering that $\phi_A=\phi_B=\phi^*$ and $f(\phi^*,\phi^*)=\phi^*$, we finally obtain
\begin{equation}\label{eqpo1}
 P_\xi(\omega) = \frac{2\phi^* \left[\gamma f_B^2 +\gamma^2+\omega^2\right]}{\omega^4+\left[(1-f_A)^2+\gamma^2+2\gamma f_B\right] \omega^2+\gamma^2(1-f_A-f_B)^2},
\end{equation}
where $f_A,f_B$ should be evaluated at the fixed point. A comparison of the results obtained from Gillespie simulations with the above result (\ref{eqpo1}) is displayed in Fig.~\ref{fig10}. Interestingly, the analytic formula predicts a peak in $P_\xi(\omega)$ for $\alpha\to\alpha_1$, which agrees with simulation results for $N_0$ large enough. This peak indicates the presence of quasi-cycles. However, when $\alpha$ approaches the critical point, the position and the shape of the peak obtained in simulations differ from that of Eq.~(\ref{eqpo1}).
Quasi-cycles which correspond to a large excursion in phase space are triggered by large fluctuations. Such fluctuations are very rare deep in the fixed-point regime, and due to their non-Gaussian nature, the induced quasi-cycles are not covered at all by the van Kampen expansion. Close to the transition region, large fluctuations are naturally  more frequent, along with them large excursions in phase space, so the discrepancy between the van Kampen results and the Gillespie simulations are more pronounced. On the other hand, the van Kampen expansion is able to predict the occurrence of quasi-cycles which are induced by small (Gaussian) fluctuations, leading to small excursions in phase space similarly to the quasi-cycles in the brusselator system \cite{galla2}. Excursions of this type are more frequent for $\alpha$-values away from the transition region; it is these quasi-cycles which induce a peak in the power spectrum at non-zero frequency both in the van Kampen expansion and the Gillespie simulations with a reasonable agreement.

\subsection{Solutions in the limit-cycle phase}
\begin{figure}
	\centering
	\psfrag{xx}{$t$} \psfrag{yy}{$\langle \xi^2(t)\rangle$} \psfrag{yy2}{$\sqrt{\frac{\langle\xi^2(t)\rangle}{N_0}}, \quad \phi_A$}
	\includegraphics*[width=\columnwidth]{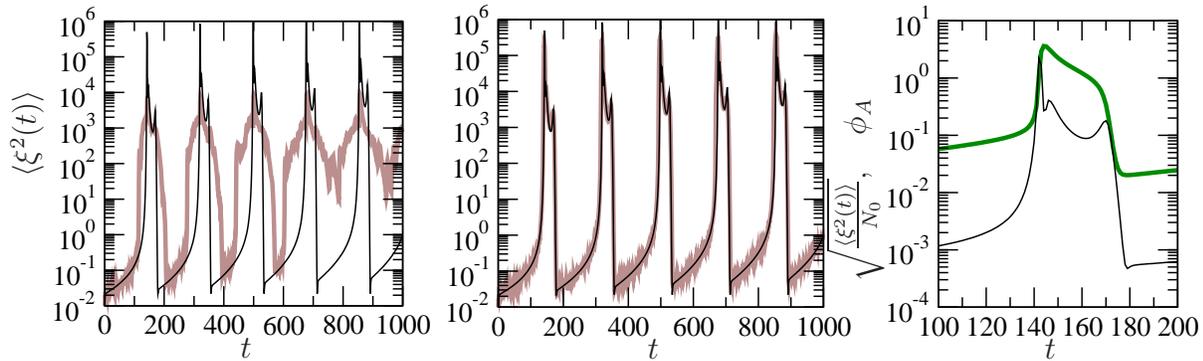} \\
	\caption{Comparison between the variance of $\xi(t)$ as function of time, obtained from 100 independent Gillespie simulations (grey line), and via the van Kampen expansion (black line), for $\alpha=50$ and $N_0=10^3$ (left) and $10^5$ (middle). The variance strongly varies along the limit cycle. Right plot: comparison of the fluctuations $\sqrt{\frac{\langle\xi^2(t)\rangle}{N_0}}$ (black line) relative to the concentration $\phi_A(t)$ (green thick line) for $N_0=10^5$.\label{fig11}}
\end{figure}

It turns out that the van Kampen size expansion works not only in the fixed-point regime, but also (to some extent as we shall see) in the limit-cycle regime, where the expansion (\ref{eqtrafo1}--\ref{eqtrafo2}) has to be carried out about the time-dependent trajectory.
We begin with calculating the variances $\langle\xi^2 (t)\rangle$, $\langle\eta^2 (t)\rangle$ along the limit cycle. Our starting point is again Eq.~(\ref{eq2mom}), but now with time-dependent functions $\phi_A(t)$, $\phi_B(t)$, $f_A(t)$, $f_B(t)$, $f(t)$, which should be determined from the limit-cycle solutions of the deterministic equations. Unlike in the fixed point, the integration cannot be done analytically in the limit-cycle regime. However, it is straightforward to integrate Eq.~(\ref{eq2mom}) together with Eqs.~(\ref{eq-det11}--\ref{eq-det12}) numerically, assuming some initial values of $\phi_A(0), \phi_B(0)$, and $\langle\xi^2 (0)\rangle = \langle\eta^2 (0)\rangle = \langle\xi(0)\eta (0)\rangle=0$. As before, the mean values $\langle\xi (t)\rangle = \langle\eta (t)\rangle = 0$ for any time $t\geq 0$. Altogether, we have five differential equations, which can be integrated by any method for solving 1st-order ordinary differential equations. In Fig.~\ref{fig11} we compare the variance $\langle\xi^2(t)\rangle$ obtained in this way for $\alpha=50$ (deeply in the limit-cycle regime) with the variance calculated from Gillespie time series according to $\langle\xi^2(t)\rangle = \langle (N_A(t)-N_0\phi_A(t))^2 \rangle/N_0$, where $\phi_A(t)$ is the deterministic solution (obtained numerically) and $\langle\dots\rangle$ denotes the average over many simulations starting from the same initial state $\phi_A(0)=0, \phi_B(0)=0.5$.
The variance oscillates with the same period $T(\alpha=50)\approx 178.1$ as the deterministic solution $\phi_A(t),\phi_B(t)$. The variance changes by many orders of magnitude along the limit cycle, with sharp peaks corresponding to fast changes in $\phi_A(t)$, see Fig. \ref{fig11}, right. For $N_0=10^5$, the quantity $\sqrt{\frac{\langle\xi^2(t)\rangle}{N_0}}/\phi_A(t)$ which measures the ratio of fluctuations to deterministic concentrations becomes of order $1$ only for a brief period of time. For smaller $N_0=10^3$, however, this ratio is larger than one for a significant part of the oscillation period, which means that the van Kampen approximation loses its applicability. As expected, the agreement between theory and simulations is thus better for large $N_0$, exactly as for the fixed-point regime, and the time period over which the van Kampen expansion applies also increases with $N_0$.

\begin{figure}
	\centering
	\psfrag{xx}{$\omega$} \psfrag{yy}{$P_{N_A}(\omega)$}
	\includegraphics*[width=14cm]{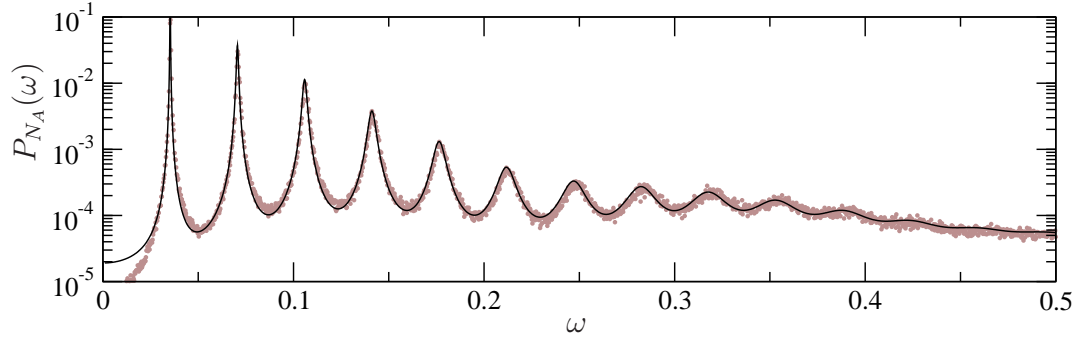}
	\caption{\label{fft_cycle}Spectrum of $N_A(t)$ for $N_0=4000, \gamma=0.01$, and $2^{26}$ Gillespie steps (grey circles), averaged over 100 realisations, compared to Eq.~(\ref{pna_final}) (black line).}
\end{figure}

Now, in principle we would like to find the autocorrelation function and the spectrum of infinite time series (that is for $t_{\rm max}\to\infty$), exactly as we did in the fixed-point regime, for finite $N_0$. This is the stationary limit of a stochastic system with non-negligible demographic fluctuations. In the appendix we shall show that (i) the van Kampen expansion is valid only for $t_{\rm max}\ll O(N_0T)$, where $T$ is the period of oscillations and so it breaks down for infinite time series, (ii) for very short $t_{\rm max}$ (a few periods of oscillations), the deterministic limit $N_0\to\infty$ is actually a reasonable approximation also for $N_0<\infty$ and the van Kampen expansion is not really needed, and (iii) in the physically relevant limit ($N_0<\infty, t_{\rm max}\to\infty$) the results of the Gillespie simulations can be analytically reproduced if we take into account that the physical time $t$ itself is a random variable, which is neglected by the van Kampen expansion. The randomness of $t$ results from the discreteness of our system which changes its state in discrete time steps of various length. This is taken into account in the Gillespie algorithm where time is incremented at each step by an exponentially distributed random number.
As explained in the appendix, we model this by assuming that variables such as $N_A,N_B$ are periodic functions of some 
 new time $x$ that plays a similar role to the number of Gillespie (time)steps. For a given fixed physical time $t$, the number of Gillespie steps varies due to random reaction times.  Similarly here, for a fixed time $t$ also $x$ varies due to stochastic increments $W(t)$, which shall reflect the fluctuations in reaction times. More precisely, $x$ is assumed to describe a Wiener process (Brownian motion) according to $dx/dt=1+W(t)$ with random numbers $W(t)$ following a Gaussian distribution with variance $\sigma^2$. It should be noticed that we assume here the increments $W(t)$ to be the only source of stochasticity, while we neglect demographic fluctuations in $N_A, N_B$ and also the possible dependence of time fluctuations on the position along the limit cycle. As explained in the Appendix, this is valid for ``spiked'' oscillations which exist in our model for $\gamma\ll 1$. We then calculate the power spectrum by averaging over these fluctuations in $x$:
\bq
	P_{N_A}(\omega) = N_0^2 \int_0^\infty dt \int_0^\infty dt' e^{i\omega(t-t')} \Big<\phi_A(x(t))\phi_A(x(t'))\Big>_x,
\eq
where the average is over different positions in the new time $x(t),x(t')$ of the Brownian motion at physical times $t,t'$, and $\phi_A(x)$ is the same deterministic solution as before, but in the time variable $x$.
For more details we refer to the appendix. The result for the power spectrum is
\bq
	P_{N_A}(\omega) \propto \sum_{n=1}^\infty  \frac{|f_n|^2 \left[n^2 \sigma^2 (n^2 \omega_0^2 + n^4 \sigma^4/4 + \omega^2)\right]}{16 n^4 \omega_0^4 +  8 n^2 \omega_0^2 (n^4 \sigma^4 - 4 \omega^2) + (n^4 \sigma^4 +  4 \omega^2)^2}, \label{pna_final}
\eq
where $f_n$ are Fourier coefficients of $\phi_A(t)$. We neglect the zero mode $\sim f_0^2 \omega^{-2}$ because it can be removed by shifting the average $N_A$ to zero. Equation (\ref{pna_final}) shows that each Fourier mode $n$, which would give a Dirac-delta peak for $N_0\to\infty$, is smeared out to a skewed Lorentzian-like function for $N_0<\infty$. In Fig.~\ref{fft_cycle} we compare this result with exact Gillespie simulations for $N_0=4000, \gamma=0.01$ and with $\sigma^2$ fitted to the data. The agreement is impressive and it suggests that we have indeed captured the main source of randomness in Gillespie simulations: the fluctuations in reaction times rather than in $N_A,N_B$.

\section{Conclusions and Outlook}
\label{sec5}
We have considered the fully stochastic version of a bistable frustrated unit which consists of a self-activating loop that activates its own repression. The unit is a simplified model of many genetic circuits. In the deterministic limit this unit shows excitable or oscillatory behaviour depending on the parameters. In our stochastic model, we have found quasi-cycles also deeply in the regions which correspond to excitable behaviour in the deterministic limit. The smaller the system size, the larger the fluctuations, and the more frequently the quasi-cycles occur. Therefore, in comparison to the deterministic version of this unit one may conclude that less fine-tuning is needed to sustain oscillations. On the other hand, quasi-cycles can still be distinguished from true cycles by looking at the autocorrelation function decay rate or the power spectrum.

It is known for systems such as deterministic FitzHugh-Nagumo units that their behaviour is much more versatile when these units are coupled with delay in the excitable regime than in the oscillatory regime \cite{boric}. It would be interesting to see whether coupling of BFUs, of which we considered a deterministic version in Ref. \cite{hmokaluza}, produces similar effects and in particular, whether the quasi-cycles of our single unit remain dynamically equivalent to genuine limit cycles of the oscillatory regime when these units are coupled. This will shed some light on the question of robustness with respect to a suitable parameter choice to maintain oscillations.

In natural genetic systems it may be challenging to disentangle oscillations which are due to internal fluctuations from those with genuine limit cycles.
What is the ``normal" mode of performance in these systems? Are quasi-cycles an adequate replacement of limit cycles even in case of delayed interactions? Most likely there is no universal answer to this question, but it will depend on the very system.

A large variety of natural genetic circuits have more complicated bistable systems coupled to negative feedback loops. Examples for such systems are the cAMP signalling system in the slime mold Dictyosthelium Discoideum \cite{51}, the embryonic division control system \cite{37,52}, or the MAPK-cascade \cite{mapk}.
In all these systems the number of reacting participants as well as the reaction events fluctuate, so that one should be aware of possible effects whose right interpretation will be in terms of inherent fluctuations.

Lastly, in Sec. 4.2 we have developed a new method of calculating power spectra of oscillating stochastic systems, simulated with the Gillespie algorithm and corresponding to biochemical reactions. The method shows that the most important source of stochasticity is the stochastic nature of the time variable $t$ and not the demographic fluctuations $\xi,\eta$ about the classical trajectory of concentrations $\phi_A,\phi_B$.
It would be very interesting to see if this remains true also for other systems like the stochastic brusselator model \cite{galla2}.

\section*{Acknowledgments}
Two of us (AG and HMO) would like to thank A. Walczak (ENS, Paris) for stimulating discussions at the beginning of this work. HN is indebted to W. Janke for discussions. BW thanks T. Galla and R. Blythe for the discussion of some details of the van Kampen size expansion, and M. R. Evans for critically reading the manuscript. Financial support from the DFG ((AG, grant no. ME-1332/17-1), (HN, grant no. JA 483/27-1)), and EPSRC (BW, grant no. EP/E030173) is gratefully acknowledged.

\section{Appendix}

\begin{figure}
	\psfrag{xx}{$t$} \psfrag{yy}{$\langle \xi^2(t)\rangle$} \psfrag{yy2}{$\sqrt{\frac{\langle\xi^2(t)\rangle}{N_0}}, \quad \phi_A$}
	\center
	\includegraphics*[width=10cm]{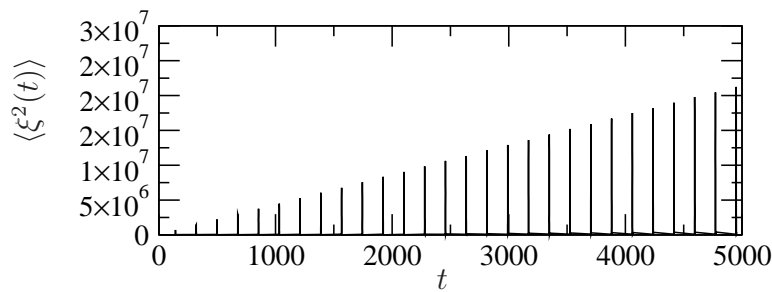} \\
	\caption{$\langle \xi^2(t)\rangle$ obtained from the van Kampen expansion grows linearly in time. Results for $\alpha=50$.\label{fig11_app}}
\end{figure}

(i) Let us first explain why the van Kampen expansion breaks down for large times. We are interested in calculating the spectrum
\bq
	P_{N_A}(\omega) = \left<|N_A(\omega)|^2\right> = \int_0^\infty dt \int_0^\infty dt' e^{i\omega(t-t')} \left<N_A(t)N_A(t')\right>. \label{eq:Ft}
\eq
Using Eq.~(\ref{eqtrafo1}) we can express $\left<N_A(t)N_A(t')\right>$ as
\bq
	\left<N_A(t)N_A(t')\right> = N_0^2 \phi_A(t)\phi_A(t') + N_0 \left<\xi(t)\xi(t')\right>. \label{eq:corr}
\eq
The above two-point correlation function has an $O(N_0^2)$ contribution from the deterministic equation, which is periodic in $t,t'$, and an $O(N_0)$ contribution from stochastic fluctuations. These fluctuations have also periodic oscillations in $t,t'$, they are however superimposed on a linear growth, see Fig.~\ref{fig11_app}.
Due to this unbounded growth, the term $N_0 \left<\xi(t)\xi(t')\right>$ eventually overcomes the term $N_0^2 \phi_A(t)\phi_A(t')$ for sufficiently large $t,t'$. This is physically unrealistic; real fluctuations cannot depart that much from the classical trajectory, because they are at most of order $\sqrt{N_0}$ for large $N_0$ unless the system is close to a transition or bifurcation point. As already noted in Ref.~\cite{galla2}, this limits the applicability of the van Kampen expansion to short time series $t_{\rm max} \ll O(N_0 T)$, and hence the method cannot accurately predict the spectrum $P_{N_A}(\omega)$ for infinite time series as we want.

\begin{figure}
	\centering
	\psfrag{xx}{$\omega$} \psfrag{yy}{$P_{N_A}(\omega)$}
	\includegraphics*[width=14cm]{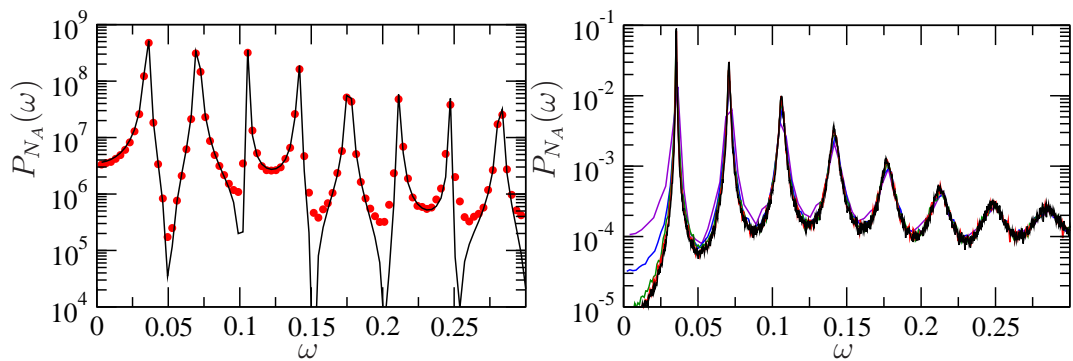}
	\caption{\label{fft_cycle_app}Left: plot of the spectrum $P_{N_A}(\omega)$ from Gillespie simulations (red points, averaged over 100 realisations for $N_0=10^5, \gamma=0.01$) and Eq.~(\ref{sp_det}) for deterministic $\phi_A(t)$ (line). Initial condition was $N_A(0)=0.0354896N_0, N_B=0.519082N_0$ which lies on the deterministic limit cycle, and the time series of length $t_{\rm max}=1904$ corresponded to $10.7$ periods of oscillation or about $2^{27}$ Gillespie steps. Right: comparison between numerically obtained spectra $P_{N_A}(\omega)$ for $N_0=4000, \gamma=0.01$ and different number of Gillespie steps (different lengths of time series) $2^{22},\dots, 2^{26}$ (curves from purple to black).}
\end{figure}

(ii) Before we proceed to an alternative method of finding $P_{N_A}(t)$ for $t_{\rm max}\to\infty$, let us consider $t_{\rm max} \approx$ a few periods $T$. As mentioned above, the expansion (\ref{eqtrafo1},\ref{eqtrafo2}) works in this regime, but its contribution to the spectrum is $\sim 1/N_0$ of the contribution from deterministic $\phi_A(t)$. We may thus hope that the deterministic spectrum $P_{\phi_A}(t)$ should be already a good approximation to the exact, finite-size $P_{N_A}(\omega)$. This is indeed true for $N_0$ large enough. To see this, let us calculate the Fourier transform of $\phi_A(t)$ for a finite period of time $t_{\rm max}$:
\bq
	\tilde{\phi}_A(\omega) = \int_0^{t_{\rm max}} e^{-i\omega t} \phi_A(t) dt .\label{phiomega}
\eq
For simplicity, let us assume that initial values $\phi_A(0),\phi_B(0)$ lie on the limit cycle. Then, $\phi_A(t)$ is periodic (with period $T$) and it can be written as a Fourier series,
\bq
	\phi_A(t)= \sum_{n=-\infty}^\infty f_n e^{i2\pi n t/T}. \label{phifn}
\eq
The coefficients $f_n$ can be determined numerically from $\phi_A(t)$. Inserting Eq.~(\ref{phifn}) into Eq.~(\ref{phiomega}) and performing integration we obtain
\bq
	\tilde{\phi}_A(\omega) = i \sum _{n=-\infty}^\infty f_n \frac{ e^{-i  (\omega - 2\pi n/T)t_{\rm max}} -1}{\omega -  2\pi n/T}, \label{tildephi}
\eq
so that the spectrum is finally given by
\bq
	P_{\phi_A}(\omega) = \left| \tilde{\phi}_A(\omega) \right|^2. \label{sp_det}
\eq
The result will generally depend on the initial condition $\phi_A(0)$. This must be taken into account when comparing with simulations which should be performed for the same initial condition as the analytical calculations: $N_A(0)=N_0\phi_A(0), N_B(0)=N_0\phi_B(0)$. In Fig.~\ref{fft_cycle_app}, left, we compare Eq.~(\ref{sp_det}) truncated to the first 30 terms in the Fourier series with $P_{N_A}(t)$ from Gillespie simulations for short $t_{\rm max} \approx 10 T$. The agreement is quite impressive, especially when taking into account that we have completely neglected stochastic effects in our calculations. Amplitudes and widths of the peaks are predicted correctly, the only discrepancy is seen in valleys between the peaks, where stochastic terms give significant contributions.

The above method cannot be applied to longer times, in particular we cannot hope that it will give correct results for $t_{\rm max}\to\infty$. In this limit, stochastic effects become significant. This is seen in Fig.~\ref{fft_cycle_app}, right, where we plotted the spectra from simulations for an increasing length of time $t_{\rm max}$. We see that the amplitudes/widths of the peaks first increase/decrease with increasing
$t_{\rm max}$, but soon they stabilise at some finite value. However, according to formulas (\ref{tildephi}) and (\ref{sp_det}), the peaks should become Dirac-delta functions for $t_{\rm max}\to\infty$. Obviously, the deterministic approach fails in this limit, and so does the van Kampen expansion. In the rest of this section we shall show how to avoid these problems and how to reproduce the spectrum measured in simulations.

(iii) We have already shown that stochastic fluctuations of $\xi(t)$ give only a small contribution to $P_{N_A}(\omega)$ for large $N_0$. Now we shall argue that the main contribution to the spectrum, which is responsible for the broadening of peaks for any finite $N_0$, comes from stochastic fluctuations of the physical time $t$. Although this is neglected in the van Kampen expansion, in stochastic systems of discrete particles $t$ is not a continuous variable but it advances in discrete steps, from one event (change in the variables $N_A,N_B$) to another one. The length of these steps depends on the position of the system in phase space. This feature is captured in Gillespie simulations, in which $t$ is a random variable, incremented with each step of the algorithm by an exponentially distributed random number.

To model the stochasticity of $t$, we shall introduce another variable $x$ which could be identified with a properly rescaled Gillespie time (number of steps) such that $x(t)$ is a Wiener process (Brownian motion) with drift and it evolves according to
\bq
	dx/dt = 1 + W(t), \label{eq:wiener}
\eq
where $W(t)$ is Gaussian noise with zero mean and variance $\sigma^2\ll 1$ which controls the strength of fluctuations (and thus it should decrease with the increase of $N_0$). In the limit $\sigma^2\to 0$ we have $t=x$ that corresponds to the deterministic limit. The variance $\sigma^2$ should in principle depend on $x$, but here we assume it to be constant and equal to the average value taken over one period of the limit cycle. This assumption simplifies calculations very much and, as it will turn out, is a very good approximation.
The reason is the following. We are mainly concerned with the case $\gamma\ll 1$, for which oscillations in $N_A(t)$ are ``spiked'', i.e., they are narrow peaks rather than smooth, sine-like oscillations. To be definite, let us consider fluctuations in the physical time which has elapsed after the system has performed one period in phase space. The main contribution to the stochasticity in this time comes from the short duration of the peak in $N_A$, where the number $N_A$ changes drastically over a few Gillespie steps (a rapid increase followed by a rapid decrease), whereas many reactions take place in the region where $N_A$ changes slowly, so that fluctuations in the number of  Gillespie steps there average out, the better, the more Gillespie steps, and the larger the system.  This means that the fluctuations in the physical time for a full period  are dominated by the temporal duration of the peak in $N_A(t)$ in relation to the number of Gillespie steps during this peak: the sharper the peak, the lower the number of Gillespie steps, the stronger  the fluctuations of the physical time when the peak occurs. For large $N_0$ the height of the peak, i.e., the maximum number of $N_A$, is  proportional to $N_0$ and stays almost the same from cycle to cycle
(corresponding to small demographic fluctuations for large $N_0$); so it is just the phase of $N_A(t)$ at which the peak occurs within the limit cycle which fluctuates from cycle to cycle. The resulting fluctuations in the physical time for a full period average out more slowly (since they are less frequent) as compared to  the demographic fluctuations and the fluctuations in the Gillespie steps outside the peak. The latter two behave similarly and decrease with increasing system size. The overall number of Gillespie steps is proportional to $N_0$. The variance $\sigma^2$ should therefore decay as $\sim 1/N_0$, as a result of summing up $O(N_0)$ exponentially distributed random variables, each of them with variance $O(1/N_0^2)$, similarly to the demographic fluctuations, so in principle both effects would give the same contribution of $O(1/\sqrt{N_0})$ to $N_A(t)$. This allows us to neglect the contribution from demographic noise and noise in Gillespie steps outside the peak,  and to concentrate on the stochasticity in time due to the peaks when they dominate the fluctuations.

We shall now assume that the evolution of $N_A=N_0\phi_A$ is fully deterministic in this new variable $x$ and that the only randomness in the evolution of $N_A(t)$ is due to the stochastic nature of $t(x)$. The spectrum $P_{N_A}(\omega)$ is then given by
\bq
	P_{N_A}(\omega) = N_0^2 \int_0^\infty dt \int_0^\infty dt' e^{i\omega(t-t')} \Big<\phi_A(x(t))\phi_A(x(t'))\Big>_x. \label{eq:Ft2}
\eq
The average $\Big<\dots\Big>_x$ is over different realizations of the Wiener process $x(t)$ and $\phi_A(x)$ is the same deterministic solution as before, but in the new variable $x$. Using the Fourier series representation of $\phi_A(x)$, we have
\bq
	\Big<\phi_A(x(t))\phi_A(x(t'))\Big>_x = \sum_{n=-\infty}^\infty \sum_{m=-\infty}^\infty f_n f_m \left<e^{i\omega_0 \left[nx(t)+mx(t')\right]}\right>_x, \label{eq:aver}
\eq
where we have introduced $\omega_0=2\pi/T$.
The average $\left<\dots\right>_x$ can be calculated using the fact that the probability distribution $P(x(t),t)$ of the Wiener process from Eq.~(\ref{eq:wiener}) is Gaussian:
\bq
	P(x,t) = \frac{1}{\sqrt{2\pi \sigma^2 t}} e^{-\frac{(x-t)^2}{2\sigma^2 t}}.
\eq
It should be evaluated for a path $x$ at two positions $t$ and $t^\prime$.
Let us focus on $t'>t$, the case $t'<t$ can be obtained by exchanging $t\leftrightarrow t'$, $n\leftrightarrow m$. The path to $x(t^\prime)$ can be split into two parts, the path to $x(t)$ and the path from $x(t)$ to $x(t^\prime)$, where the difference $x(t^\prime)-x(t)$ is a new random variable $y(t^\prime-t)$ that follows a Wiener process (\ref{eq:wiener}), but during the period $t'-t$. This possible splitting is an expression of the fact that a Wiener process has no memory to the history of how the system reached the point $x(t)$.
We have
\ba
	\left<e^{i\omega_0 \left[nx(t)+mx(t')\right]}\right>_x &=& \left<e^{i\omega_0 \left[(n+m)x(t)+my(t'-t)\right]}\right>_{x,y} \nonumber \\
	&=& \Big<e^{i\omega_0 (n+m)x(t)}\Big>_{x} \left<e^{i\omega_0 my(t'-t)}\right>_{y}, \label{avxy}
\ea
where $y(t'-t)$ has the probability distribution
\bq
	P(y,t'-t) = \frac{1}{\sqrt{2\pi \sigma^2 (t'-t)}} e^{-\frac{\left(y-(t'-t)\right)^2}{2\sigma^2 (t'-t)}}.
\eq
The averages in Eq.~(\ref{avxy}) can be then written as Gaussian integrals:
\ba
	\Big<e^{i\omega_0 (n+m)x(t)}\Big>_{x} &\left<e^{i\omega_0 my(t'-t)}\right>_{y} = \frac{1}{\sqrt{2\pi \sigma^2 t}} \int_{-\infty}^\infty  e^{-\frac{(x-t)^2}{2\sigma^2 t}+i\omega_0 (n+m)x} dx \nonumber \\
	&\times \frac{1}{\sqrt{2\pi \sigma^2 (t'-t)}}  \int_{-\infty}^\infty e^{-\frac{\left(y-(t'-t)\right)^2}{2\sigma^2 (t'-t)}+i\omega_0 my} dy,
\ea
and they can be easily calculated. Inserting the result back into Eqs.~(\ref{eq:aver}) and (\ref{eq:Ft2}) and performing the integrals over $t,t'$ we obtain (after some tedious calculations) the spectrum (\ref{pna_final}).


\section*{References}

\begin{thebibliography}{100}
\bibitem{newman} McKane A J and Newman T J, {\it Predator-prey cycles from resonant amplification of demographic stochasticity}, 2005 Phys. Rev. Lett. {\bf 94} 218102.
\bibitem{traulsen} Traulsen A, Claussen J C, and Hauert C, {\em Coevolutionary dynamics: from finite to infinite populations}, 2005 Phys. Rev. Lett. {\bf 95} 238701.
\bibitem{golden1} Butler T, and Goldenfeld N, {\em Robust ecological pattern formation induced by demographic noise}, 2009 Phys. Rev. E {\bf 80} 030902.
\bibitem{alex1} Schultz D, Onuchic J N, and Wolynes P G, {\em Understanding stochastic simulations of the smallest genetic networks }, 2007 J. Chem. Phys. {\bf 126} 245102.
\bibitem{alex2} Hornos J E, Schultz D, Innocentini G C P, Wang J, Walczak A M, Onuchic J N, and Wolynes P G, {\em Self-regulating gene: An exact solution}, 2005 Phys. Rev. E {\bf 72} 051907.
\bibitem{alex3} Schultz D, Walczak A M, Onuchic J N, and Wolynes P G, {\em Extinction and resurrection in gene networks}, 2008 Proc. Natl. Acad. Sci. USA {\bf 105} 19165.
\bibitem{yoda} Yoda M, Ushikubo T, Inoue W, and Sasai M, {\em Roles of noise in single and coupled multiple genetic oscillators}, 2007 J. Chem. Phys. {\bf 126} 115101.
\bibitem{loinger} Loinger A and Biham O, {\em Stochastic simulations of the repressilator circuit}, 2007 Phys. Rev. E {\bf 76} 051917.
\bibitem{galla1} Galla T, {\em Intrinsic fluctuations in stochastic delay systems: Theoretical description and application to a simple model of gene regulation}, 2009 Phys. Rev. E {\bf 80} 021909.
\bibitem{sandeep} Krishna S, Semsey S, and Jensen M, {\em Frustrated bistability as a means to engineer oscillations in biological systems}, 2009 Phys. Biol. {\bf 6} 036009.
\bibitem{fitzhugh} FitzHugh R, {\em Impulses and physiological states in theoretical models of nerve membrane}, 1961 Biophys. J. {\bf 1} 445.
\bibitem{nagumo} Nagumo J S, Arimoto S, and Yoshizawa S, {\em An Active Pulse Transmission Line Simulating Nerve Axon }, 1962 Proc. IRE {\bf 50} 2061.
\bibitem{hmokaluza} Kaluza P, and Meyer-Ortmanns H, {\em On the role of frustration in excitable systems}, 2010 Chaos {\bf 20} 043111-1-11.
\bibitem{hwa} Scott M, Hwa T, and Ingalls B, {\em Deterministic characterization of stochastic genetic circuits}, 2007 Proc. Natl. Acad. Sci. {\bf 104} 7402.
\bibitem{leibler} Vilar J M G, Kueh H Y, Barkai N, and Leibler S, {\em Mechanisms of noise-resistance in genetic oscillators}, 2002 Proc. Natl. Acad. Sci. {\bf 99} 5988.
\bibitem{golden2} Butler T, and Goldenfeld N, {\em Fluctuation-driven Turing patterns}, 2011 Phys. Rev. E {\bf 84} 011112.
\bibitem{turing} Turing A M, {\em The chemical basis of morphogenesis }, 1952 Philos. Trans. R. Soc. London B {\bf 237} 37.
\bibitem{galla0} Bladon A J, Galla T, and McKane A J, {\em Evolutionary dynamics, intrinsic noise, and cycles of cooperation}, 2010 Phys. Rev. E {\bf 81} 066122.
\bibitem{bressloff} Bressloff P C, Cowan J D, Golubitsky M, Thomas P J, and Wiener M C, {\em What geometric visual hallucinations tell us about the visual cortex}, 2002 Neural Computation {\bf 14} 473.
\bibitem{galla2} Boland R P, Galla T, and McKane A J, {\em How limit cycles and quasi-cycles are related in systems with intrinsic noise}, 2008 J. Stat. Mech. P09001.
\bibitem{pomerening} Pomerening J R, Kim S Y, and Ferrell J E Jr., {\em Systems-level dissection of the cell-cycle oscillator: bypassing positive feedback produces damped oscillations}, 2005 Cell {\bf 122} 565.
\bibitem{guantes} Guantes R and Poyatos J F, {\em Dynamical principles of two-component genetic oscillators}, 2006 PLoS Comp. Biol. {\bf 2} 0188.
\bibitem{strogatz} Strogatz S H, {\it Nonlinear Dynamics and Chaos: With Applications to Physics, Biology, Chemistry and Engineering}, 1994 Westview Press, Perseus Books Publishing LLC, Cambridge, MA.
\bibitem{ingolia} Ingolia N T and Murray A W, {\em The ups and downs of modeling the cell cycle}, 2004 Current Biology {\bf 14} R771.
\bibitem{forthcoming} Nagel H, Labavic D, Janke W, and Meyer-Ortmanns H, in preparation.
\bibitem{gillespie} Gillespie D T, {\em Exact stochastic simulation of coupled chemical reactions}, 1977 J. Phys. Chem. {\bf 81} 2340.
\bibitem{bratsun} Bratsun D, Volfson D, Tsimring L S, and Hasty J, {\em Delay-induced stochastic oscillations in gene regulation}, 2005, Proc. Natl. Acad. Sci. USA {\bf 102} 14593.
\bibitem{vankampen} Van Kampen N G, {\it Stochastic Processes in Physics and Chemistry}, 2005 Elsevier, Amsterdam, The Netherlands.
\bibitem{boric} Buri$\acute{c}$ N, and Todorovi$\acute{c}$ D, {\em Dynamics of FitzHugh-Nagumo excitable systems with delayed coupling}, 2003 Phys. Rev. E {\bf 67} 066222.
\bibitem{51} Martiel J, and Goldbeter A, {\em A model based on receptor desensitization for cyclic AMP signaling in Dictyostelium cells}, 1987 Biophys. J. {\bf 52} 807.
\bibitem{37} Novak B, and Tyson J J, {\em Numerical analysis of a comprehensive model of M-phase control in Xenopus oocyte extracts and intact embryos}, 1993 J. Cell Sci. {\bf 106} 1153.
\bibitem{52} Tyson J J, {\em Modeling the cell division cycle: cdc2 and cyclin interactions}, 1991 Proc. Natl. Acad. Sci. USA {\bf 88} 7328.
\bibitem{mapk} Qiao L, Nachbar R B, Kevrekidis I G, and Shvartsman S Y, {\em Bistability and oscillations in the Huang-Ferrell model of MAPK signaling}, 2007 PLoS Computational Biology {\bf 3} e184.
\end{thebibliography}
\end{document}